\documentclass[twocolumn,showpacs,preprintnumbers,amsmath,amssymb]{revtex4}
\usepackage{epsfig}
\usepackage{graphicx}
\usepackage{dcolumn}
\usepackage{bm}
\usepackage{amsthm}
\usepackage{amsfonts}
\usepackage{color} 

\newcommand{\ket}[1]{\mbox{$ | #1 \rangle $}}
\newcommand{\bra}[1]{\mbox{$ \langle #1 | $}}

\begin{document}

\title{Passive decoy state quantum key distribution with practical light sources}
\author{Marcos Curty$^1$, Xiongfeng Ma$^2$, Bing Qi$^3$, and
Tobias Moroder$^{2,4,5}$}
\affiliation{$^1$ ETSI Telecomunicaci\'on, Department of Signal Theory and Communications, University of Vigo, 
Campus Universitario, E-36310 Vigo (Pontevedra), Spain \\
$^2$ Institute for Quantum Computing \& Department of Physics and Astronomy, University of Waterloo, N2L 3G1 
Waterloo, Ontario, Canada \\
$^3$ Center for Quantum Information and Quantum Control, Department of Physics and Department of 
Electrical \& Computer Engineering, University of Toronto, M5S 3G4 Toronto, Ontario, Canada \\
$^4$ Quantum Information Theory Group, Institute of Theoretical Physics I, University of Erlangen-N\"urnberg, 
91058 Erlangen, Germany \\
$^5$ Max Planck Institute for the Science of Light,
91058 Erlangen, Germany}

\begin{abstract}
Decoy states have been proven to be a very useful method for significantly enhancing 
the performance of quantum key distribution systems with practical light sources. 
While active
modulation of the intensity of the laser pulses is an effective way of preparing
decoy states in principle, in practice passive preparation might be desirable in 
some scenarios.
Typical passive schemes involve parametric down-conversion. More recently, it has been 
shown that phase randomized weak coherent pulses (WCP) can also be used for the same
purpose [M. Curty {\it et al.}, Opt. Lett. {\bf 34}, 3238 (2009).] This proposal 
requires only linear optics together with a simple
threshold photon detector, which shows the practical feasibility of the method.
Most importantly, the resulting secret key rate is comparable to the one delivered by an active
decoy state setup with an infinite number of decoy settings. 
In this paper we extend these results, now showing specifically the 
analysis for other practical scenarios with different light sources and photo-detectors. 
In particular, we consider sources emitting thermal states, phase randomized 
WCP, and strong coherent light in 
combination with several types of photo-detectors, like, for instance, threshold photon detectors, photon number 
resolving detectors, and classical photo-detectors. Our analysis includes as well the 
effect that detection inefficiencies and noise in the form of dark counts shown by current threshold detectors 
might have on the final secret ket rate. Moreover, we provide estimations on the effects that 
statistical fluctuations due to a finite data size can have in practical implementations.
\end{abstract}

\maketitle

\section{Introduction}

Quantum key distribution (QKD) is the first quantum information task that reaches the commercial market 
to offer efficient and user-friendly cryptographic systems providing an unprecedented level of security
\cite{comp}. It allows two distant
parties (typically called Alice and Bob)
to establish a secure secret key despite
the computational and technological power of an eavesdropper (Eve), who interferes with the signals \cite{qkd}. 
This secret key is the essential ingredient of the one-time-pad or Vernam cipher \cite{vernam}, the only
known encryption method that 
can deliver information-theoretic secure communications.

Practical implementations of QKD 
are usually based on the transmission of phase randomized weak coherent pulses (WCP) 
with typical average photon number of $0.1$ or higher \cite{wcp}. 
These states can be easily prepared using only standard semiconductor lasers and calibrated 
attenuators. The main drawback of these systems, however, arises from the fact that some 
signals may contain 
more than one photon prepared in the same quantum state. 
When this effect is combined with the considerable attenuation introduced 
by the quantum channel (about $0.2$ dB/km), it opens an important security loophole. 
Eve can perform, for instance, the so-called {\it Photon Number Splitting} attack on the multi-photon pulses 
\cite{pns}. 
This attack provides her with full information about the part of the key generated with the multi-photon 
signals, without causing any disturbance in the signal polarization. 
As a result, it turns out that the standard BB84 protocol \cite{bb84} with phase randomized 
WCP can deliver a key generation rate of order
 $O(\eta^2)$, where $\eta$ denotes the transmission efficiency of the quantum channel \cite{opt,gllp}. 
This poor performance contrasts 
with the one expected from
a QKD scheme using
a single photon source, where the key generation rate scales linearly with $\eta$.

A significant improvement of the achievable secret key rate 
can be obtained if the original hardware is 
slightly modified. For instance, one can use the so-called decoy state method 
\cite{decoy1,decoy2,decoy3,estimation}, 
which can basically 
reach the performance of single photon sources. The essential idea 
behind decoy state QKD with phase randomized WCP is quite simple:
Alice varies, independently and randomly, the mean photon number 
of each signal state she sends to Bob by employing different intensity settings. 
This is typically realized by means of a variable optical attenuator (VOA)
together with a random number generator. 
Eve does not know 
a priori the mean photon number of each signal state sent by Alice. 
This means that 
her eavesdropping strategy can only depend on the actual photon number of these signals, but not on 
the particular intensity setting used to generate them. From the measurement results corresponding to 
different intensity settings, the legitimate users can obtain a better estimation 
of the behavior of the quantum channel.
This fact translates into an enhancement of the resulting 
secret key rate.
The decoy state technique has been successfully implemented in several recent experiments \cite{decoy_e}, 
which show the practical feasibility of this method.

While active modulation of the intensity of the pulses suffices to perform decoy state QKD in principle, 
in practice passive preparation might be desirable in some scenarios. 
For instance, in those experimental setups operating at high transmission rates. Passive schemes 
might also be more resistant to side channel attacks than active systems. For example, if 
the VOA which changes the intensity of Alice's pulses is not properly designed, it may happen
that some physical parameters of the pulses emitted by the sender depend on the particular setting selected. 
This fact could open a security loophole in the active schemes. 

Known passive schemes rely typically on the use of 
a parametric down-conversion (PDC) source 
together with a photon detector \cite{mauerer,mauerer2,mauerer3}. The main idea behind these
proposals comes from the 
photon number correlations that exist between the two output modes of a PDC source. 
By measuring the photon number distribution of one output mode it is 
possible to infer the photon number statistics of the other mode. In particular, Ref.~\cite{mauerer} considers the 
case where Alice measures one of the output modes by means of  
a time multiplexed detector (TMD) which provides photon number 
resolution capabilities \cite{tmd}; Ref.~\cite{mauerer2} analyzes the scenario where the 
detector used by Alice is just a simple threshold detector, while the authors of Ref.~\cite{mauerer3} generalize 
the ideas introduced by Mauerer {\it et al.} in Ref.~\cite{mauerer} to QKD setups using triggered PDC sources. 
All these schemes nearly reach the performance of a single photon source. 

More recently, it has been 
shown that phase randomized WCP can also be used for the same
purpose \cite{mcurty_opt}. That is, 
one does not need a non-linear optics network preparing entangled states. 
The crucial requirement of a passive decoy state setup is to obtain correlations 
between the photon number statistics of different signals; hence it is sufficient that these correlations are classical.
The main contribution of Ref.~\cite{mcurty_opt}
is rather simple: When
two phase randomized coherent states interfere at a beam splitter (BS), the photon number statistics of the outcome signals are 
classically correlated. This effect contrasts with the one expected from the 
interference of two pure coherent states with fixed phase relation at a BS. In this last case, it is well known that  
the photon number statistics of the outcome signals is just the 
product of two Poissonian distributions. Now the idea is similar to that of 
Refs.~\cite{mauerer,mauerer2,mauerer3}: By measuring one of 
the two outcome signals of the BS, the conditional photon number distribution of the other signal varies depending 
on the result obtained \cite{mcurty_opt}. In the asymptotic limit of an infinite long experiment, 
it turns out that 
the secret key rate provided by such a passive scheme 
is similar to the one delivered by an active
decoy state setup with infinite decoy settings \cite{mcurty_opt}.
A similar result can also be obtained when Alice uses
heralded single-photon sources showing non-Poissonian photon number statistics \cite{masato}. 

In this paper we extend the results presented in Ref.~\cite{mcurty_opt}, 
now showing specifically the 
analysis for other practical scenarios with different light sources and photo-detectors. 
In particular, we consider sources 
emitting thermal states and phase randomized 
WCP in combination with threshold detectors and photon number 
resolving (PNR) detectors. In the case of threshold detectors, we
include as well the 
effect that detection inefficiencies and dark counts present in current measurement devices 
might have on the final secret ket rate. For simplicity, 
these measurement imperfections  were not considered in Ref.~\cite{mcurty_opt}.
On the other hand, PNR detectors allows us 
to obtain ultimate 
lower bounds on the maximal performance that can be expected at all from this kind of passive setups. 
We also present a passive scheme that employs
strong coherent light and does not require the use of  
single photon detectors, but it can operate with a simpler
classical photo-detector. This fact makes this setup specially interesting
from an experimental point of view. Finally, we provide an estimation on the effects 
that statistical fluctuations due to a finite data size can have in practical implementations. 

The paper is organized as follows. In Sec.~\ref{sec_one} we review very briefly the concept of decoy state QKD. 
Next, in Sec.~\ref{sec_model} we present a simple model to characterize the behavior of a typical quantum 
channel. This model will be 
relevant later on, when we evaluate 
the performance of the different passive schemes that we present 
in the following sections. Our starting point is the basic passive decoy state 
setup introduced in Ref.~\cite{mcurty_opt}. 
This scheme is explained very briefly in 
Sec.~\ref{sec_tres}. Then, 
in Sec.~\ref{sec_thermal} we analyze its security 
when Alice uses a source of thermal light. 
Sec.~\ref{sec_weak} and Sec.~\ref{sec_strong}
consider the case where Alice
employs 
a source of coherent light. First, 
Sec.~\ref{sec_weak} investigates the scenario where the states prepared by Alice are phase randomized WCP.
Then, Sec.~\ref{sec_strong} presents a passive decoy state scheme that uses 
strong coherent light. In Sec.~\ref{fluc} we discuss the effects of statistical fluctuations. 
Finally, Sec.~\ref{conc} concludes the paper with a summary. 

\section{Decoy state QKD}\label{sec_one}

In decoy state QKD Alice prepares mixtures of Fock states with different photon number statistics and sends these states  
to Bob \cite{decoy1,decoy2,decoy3, estimation}. 
The photon number distribution of each signal state is chosen, independently and at random, from a set of 
possible predetermined settings. Let $p_n^l$ denote the conditional probability that a signal state 
prepared by Alice contains $n$ photons given that she selected setting $l$, with $l\in\{0,\ldots,m\}$. 
For instance, 
if Alice employs a source of phase randomized WCP then 
$p_n^l=e^{-\mu_l}\mu_l^n/n!$, and she varies the mean photon number (intensity) $\mu_l$ of 
each signal. Assuming that Alice has choosen
setting $l$, such states can be described as
\begin{equation}
\rho^l=\sum_{n=0}^\infty p_n^l \ket{n}\bra{n},
\end{equation}
where $\ket{n}$ denote Fock states with $n$ photons.

The gain $Q^l$ corresponding to setting $l$, {\it i.e.}, the probability that 
Bob obtains a click in his measurement apparatus 
when Alice sends him a signal state prepared with setting $l$, can be written as
\begin{equation}\label{gain_rome}
Q^{l}=\sum_{n=0}^\infty p^{l}_{n}Y_n, 
\end{equation}
where $Y_n$ denotes the yield of an $n$-photon signal, {\it i.e.},
the conditional 
probability of a detection event on Bob's side given that Alice transmitted an $n$-photon state. Similarly, 
the quantum bit error rate (QBER) 
associated to setting $l$, that we shall denote as $E^l$, is given by
\begin{equation}\label{rome3}
Q^{l}E^{l}=\sum_{n=0}^\infty p^{l}_{n}Y_ne_n,
\end{equation}
with $e_n$ representing the error rate of an $n$-photon signal. 

Now the main idea of decoy state QKD is very simple. From the observed data $Q^l$ and $E^l$, 
together with the 
knowledge of the photon number 
distributions $p_n^l$, Alice and Bob can estimate the value 
of the unknown parameters $Y_n$ and $e_n$ just by 
solving the set of linear equations given by 
Eqs.~(\ref{gain_rome})-(\ref{rome3}).
For instance, in the general 
scenario where Alice employs an infinite number of possible decoy settings then she can 
estimate any finite number of parameters $Y_n$ and $e_n$ with arbitrary precision. 
On the other hand, if Alice and Bob are only interested in the 
value of a few probabilities (typically $Y_0$, $Y_1$, and $e_1$), then they can estimate them by means of only a 
few different decoy settings \cite{decoy2,decoy3,estimation}.

In this paper we shall consider that Alice and Bob treat each decoy setting separately, and they distill secret key from 
all of them. We use the security analysis presented in Ref.~\cite{decoy2}, which combines the results provided
by Gottesman-Lo-L\"utkenhaus-Preskill (GLLP) in Ref.~\cite{gllp} (see also Ref.~\cite{lo_qic}) with the decoy state 
method. Specifically, the secret key rate formula can be written as
\begin{equation}\label{key_rate}
R\geq{}\sum_{l=0}^m\textrm{max}\{R^{l},0\},
\end{equation}
where $R^{l}$ satisfies 
\begin{equation}\label{ind_kr_new}
R^{l}\geq{}q\{-Q^{l}f(E^{l})H(E^{l})+p^{l}_{1}Y_1[1-H(e_1)]+p^{l}_{0}Y_0\}.
\end{equation}
The parameter $q$ is the efficiency of the protocol ($q=1/2$ for the standard BB84
protocol \cite{bb84}, and $q\approx{}1$ for its efficient version \cite{eff_bb84});
$f(E^{l})$ is the efficiency of the error correction protocol as a function of the error rate $E^{l}$
\cite{eff_error}, typically $f(E^{l})\geq{}1$ with Shannon limit $f(E^{l})=1$;   
$e_1$ denotes the single photon error rate; 
$H(x)=-x\log_2{(x)}-(1-x)\log_2{(1-x)}$ is the binary 
Shannon entropy function. 

To apply the secret key rate formula given by Eq.~(\ref{ind_kr_new}) one needs to solve 
Eqs.~(\ref{gain_rome})-(\ref{rome3}) in order to estimate 
the quantities $Y_0$, $Y_1$, and $e_1$. 
For that, we shall use the procedure proposed in Ref.~\cite{estimation}.
This method requires that the probabilities $p_n^l$ satisfy certain conditions. 
It is important to 
emphasize, however, that the estimation 
technique presented in Ref.~\cite{estimation} only constitutes a possible example of a finite 
setting estimation procedure and no optimality statement is given. In principle, 
many other estimation methods are also available for this purpose, like, for instance, linear 
programming tools \cite{linear}, which might result in a sharper, or for the purpose of QKD
better, bounds on the considered probabilities. 

\section{Channel model}\label{sec_model}

In this section we present a simple model to describe the behavior of a typical quantum channel.
This model will be relevant later on, 
when we evaluate the performance of the 
passive decoy state setups that we present in the following sections. In particular, we shall consider the channel model 
used in Refs.~\cite{decoy2,estimation}. This model reproduces a normal behavior 
of a quantum channel, {\it i.e.}, in the absence of eavesdropping.  
Note, however, that 
the results presented in this paper can also be
applied to any other quantum channel, as they only depend on the observed
gains $Q^l$ and error rates $E^l$. 

\subsection{Yield}

There are two main factors that contribute to the 
yield of an $n$-photon signal: The background rate $Y_0$, and 
the signal states sent by Alice. Usually $Y_0$ is, to a good approximation, independent of 
the signal detection. This parameter depends mainly on the dark 
count rate of Bob's detection apparatus, together with other background 
contributions like, for instance, stray light coming from timing pulses which are not 
completely filtered out in reception. In the scenario considered, the yields $Y_n$ can be expressed as
\cite{decoy2,estimation}
\begin{equation}\label{yield_new}
Y_n=1-(1-Y_0)(1-\eta_{\rm sys})^n,
\end{equation}
where $\eta_{\rm sys}$ represents the overall 
transmittance of the system. This quantity can be written as
\begin{equation}
\eta_{\rm sys}=\eta_{\rm channel}\eta_{\rm Bob},
\end{equation}
where $\eta_{\rm channel}$ is the transmittance of the quantum channel, and
$\eta_{\rm Bob}$ denotes the overall transmittance of Bob's detection apparatus. That is, $\eta_{\rm Bob}$ includes 
the transmittance of any optical component within Bob's measurement device and the 
detector efficiency. The parameter $\eta_{\rm channel}$
can be related with a transmission distance $d$ measured in km for 
the given QKD scheme as 
\begin{equation}
\eta_{\rm channel}=10^{-\frac{\alpha{}d}{10}},
\end{equation}
where $\alpha$ represents 
the loss coefficient of the channel ({\it e.g.}, an optical fiber) measured in dB/km. 

\subsection{Quantum bit error rate}

The $n$-photon error rate $e_n$ is given by \cite{decoy2,estimation}
\begin{equation}\label{qber_new}
e_n=\frac{Y_0e_0+(Y_n-Y_0)e_d}{Y_n},
\end{equation}
where $e_{d}$ is the probability that a signal hits the wrong detector on Bob's side due to the 
misalignment in the quantum channel and in his detection setup. For simplicity, 
here we assume that $e_d$ is a constant independent of the distance. Moreover, from now on 
we shall consider that the background is random, {\it i.e.}, $e_0=1/2$. 

\section{Passive decoy state QKD setup}\label{sec_tres}

\begin{figure}
\begin{center}
\includegraphics[angle=0,scale=0.62]{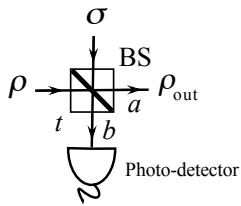}
\end{center}
\caption{Basic setup of a passive decoy state QKD scheme: Interference of two Fock diagonal states, $\rho$ and $\sigma$, 
at a beam splitter (BS) 
of transmittance $t$; $a$ and $b$ represent the two output modes.
\label{figure_general}}
\end{figure}

The basic setup is rather simple \cite{mcurty_opt}. 
It is illustrated in Fig.~\ref{figure_general}. Suppose two Fock diagonal states
\begin{eqnarray} 
\rho&=&\sum_{n=0}^\infty p_n \ket{n}\bra{n}, \nonumber \\ 
\sigma&=&\sum_{n=0}^\infty r_n \ket{n}\bra{n}, 
\end{eqnarray}
interfere at a BS of transmittance t. 
If the probabilities $p_n$ and $r_n$ are properly selected, then it turns out that the photon number distributions 
of the two outcome signals can be classically correlated. By measuring the signal state in mode $b$, therefore, the 
conditional photon number statistics of the signal state in mode $a$ vary depending on the result obtained. 

In the following sections we analyze the setup represented in Fig.~\ref{figure_general} for different 
light sources and photo-detectors. We start by considering a simple source of thermal states. Afterwards, we investigate 
more practical sources of coherent light.

\section{Thermal light}\label{sec_thermal}

Suppose that the signal state 
$\rho$ which appears in Fig.~\ref{figure_general}
is a thermal state of mean photon number $\mu$. Such state can be written as
\begin{equation}\label{rome_ns}
\rho=\frac{1}{1+\mu}\sum_{n=0}^{\infty} \Big(\frac{\mu}{1+\mu}\Big)^n \ket{n}\bra{n},
\end{equation}
and let $\sigma$ be a vacuum state. In 
this scenario, 
the joint probability of having $n$ photons in output mode $a$ and $m$ photons in
output mode $b$ (see Fig.~\ref{figure_general}) has the form 
\begin{equation}\label{prob_th}
p_{n,m}=\frac{1}{1+\mu}\binom{n+m}{m} \Big(\frac{\mu}{1+\mu}\Big)^{n+m} t^n(1-t)^m.
\end{equation}
That is, depending on the result of Alice's measurement in mode $b$, 
the conditional photon number distribution of the signals in mode $a$ 
varies. 

In particular, we have that whenever Alice ignores the result of her measurement,
the total probability of finding  
$n$ photons in mode $a$ can be expressed as 
\begin{equation}\label{help_thermal}
p^t_{n}=\sum_{m=0}^{\infty}p_{n,m}=\frac{1}{1+\mu{}t}\Big(\frac{\mu{}t}{1+\mu{}t}\Big)^n.
\end{equation}  
Next, we consider the case where Alice uses a threshold detector to measure mode $b$. 

\subsection{Threshold detector}\label{thermal_threshold}

Such a detector can be characterized by a positive operator value measure (POVM) 
which contains two elements, $F_{\rm{vac}}$ and $F_{\rm click}$, given by
\cite{detector}
\begin{eqnarray}\label{rome_final}
F_{\rm{vac}}&=&(1-\epsilon)\sum_{n=0}^{\infty} (1-\eta_{\rm d})^n \ket{n}\bra{n}, \nonumber \\
F_{\rm click}&=&\openone-F_{\rm{vac}}.
\end{eqnarray}
The parameter $\eta_{\rm d}$
denotes the detection efficiency of the 
detector, and $\epsilon$ represents its probability of having a dark count. Eq.~(\ref{rome_final}) 
assumes that $\epsilon$ is, to a good 
approximation, independent of the incoming signals.
The outcome of $F_{\rm{vac}}$ corresponds to ``no click'' in the detector, while the operator 
$F_{\rm click}$ gives precisely one detection ``click'', which means at least one photon is 
detected.

\begin{figure}
\begin{center}
\includegraphics[angle=0,scale=0.72]{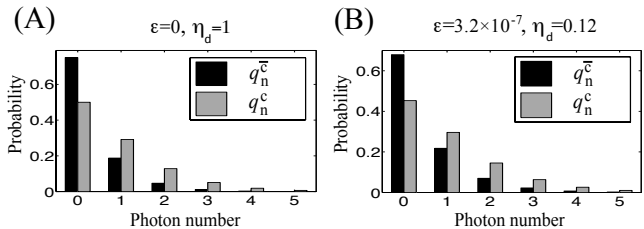}
\end{center}
\caption{
Conditional photon number distribution in mode $a$ (see Fig.~\ref{figure_general}): 
$q^{\bar{c}}_{n}$ (black) versus $q^{c}_{n}$ (grey) when
$\rho$ is given by Eq.~(\ref{rome_ns}), and $\sigma$ is a vacuum state.
We use $\mu=1$ and $t=1/2$, and 
we study two situations: (A) A perfect threshold photon detector, {\it i.e.},  
$\epsilon=0$ and $\eta_{\rm d}=1$, and (B) $\epsilon=3.2\times{}10^{-7}$ and $\eta_{\rm d}=0.12$. 
These last data correspond to the experiment reported by 
Gobby {\it et al.} in Ref.~\cite{gys}. \label{figure2_thermal}}
\end{figure}

The joint probability for seeing $n$
photons in mode $a$ and no click in the threshold detector, which we shall denote as 
$p^{\bar{c}}_n$, has the form
\begin{equation}\label{pnc_thermal_imp}
p^{\bar{c}}_n=(1-\epsilon)\sum_{m=0}^\infty (1-\eta_{\rm d})^mp_{n,m}
=\frac{(1-\epsilon)}{r}\Big(\frac{\mu{}t}{r}\Big)^n,
\end{equation}
with the parameter $r$ given by
\begin{equation}\label{parameter_r}
r=1+\mu[t+(1-t)\eta_{\rm d}].
\end{equation}
If the detector produces a click, 
the joint probability of finding $n$ photons in mode $a$ 
is given by 
\begin{equation}\label{rome_tues}
p^{c}_{n}=p^{t}_{n}-p^{\bar{c}}_{n}.
\end{equation}
Figure~\ref{figure2_thermal} shows the conditional photon number statistics
of the outcome signal in mode $a$ depending on the result of the threshold detector
(click and not click): $q^{c}_{n}=p^{c}_{n}/(1-N_{\rm th})$ and $q^{\bar{c}}_n=p^{\bar{c}}_n/N_{\rm th}$, with
\begin{equation}
N_{\rm th}=\sum_{n=0}^{\infty}p^{\bar{c}}_{n}=\frac{1-\epsilon}{1+\mu\eta_{\rm d}(1-t)}.
\end{equation}

\subsection{Lower bound on the secret key rate}\label{rome_cans}

We consider that Alice and Bob distill secret key both from
click and no click events. The calculations to estimate the yields $Y_0$ and $Y_1$, together with 
the single photon error rate $e_1$, are included 
in Appendix~\ref{ap_vico1}. 

For simulation purposes we use the 
channel model described in Sec.~\ref{sec_model}. 
After substituting 
Eqs.~(\ref{yield_new})-(\ref{qber_new}) into the gain and QBER formulas we obtain that the parameters 
$Q^{\bar c}$, $E^{\bar c}$, $Q^t$, and $E^t$
can be written as
\begin{eqnarray}
Q^{\bar{c}}&=&N_{\rm th}-\frac{(1-\epsilon)(1-Y_0)}{r-(1-\eta_{\rm sys})\mu{}t}, \nonumber \\
Q^{\bar{c}}E^{\bar{c}}&=&(e_0-e_d)Y_0N_{\rm th}+e_dQ^{\bar{c}}, \nonumber \\
Q^{t}&=&\frac{Y_0+\mu{}t\eta_{\rm sys}}{1+\mu{}t\eta_{\rm sys}}, \nonumber \\
Q^{t}E^{t}&=&(e_0-e_d)Y_0+e_dQ^{t},
\end{eqnarray}
where  $Q^{c}=Q^{t}-Q^{\bar{c}}$ and $Q^{c}E^{c}=Q^{t}E^{t}-Q^{\bar{c}}E^{\bar{c}}$.

\begin{figure}
\begin{center}
\includegraphics[angle=0,scale=0.35]{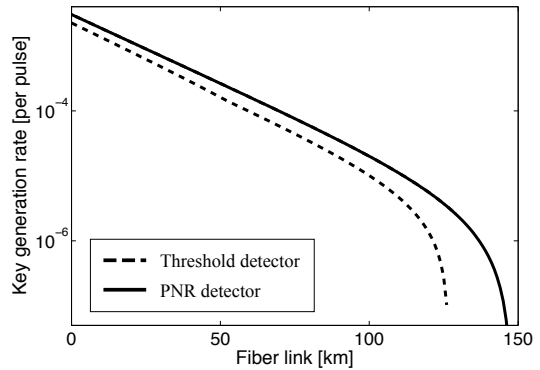}
\end{center}
\caption{
Lower bound on the secret key rate $R$ given by Eq.~(\ref{key_rate}) 
in logarithmic scale 
for the passive decoy state setup illustrated in Fig.~\ref{figure_general} 
with two intensity settings. 
The signal state $\rho$ is given by Eq.~(\ref{rome_ns}), and $\sigma$ is a vacuum state.
We consider two possible scenarios: (A) A perfect threshold detector, 
{\it i.e.}, $\epsilon=0$ and $\eta_{\rm d}=1$, and
(B) $\epsilon=3.2\times{}10^{-7}$ and $\eta_{\rm d}=0.12$ \cite{gys}.
Both cases provide approximately the same final key rate and they cannot 
be distinguished with the resolution of this figure (dashed line). 
The solid line
represents a lower bound on $R$ when Alice employs a PNR detector 
instead of a threshold detector (see Appendix~\ref{ap_pnr}). 
\label{figure3_thermal_combined}}
\end{figure}

The resulting lower bound on the secret key rate is illustrated in Fig.~\ref{figure3_thermal_combined} (dashed line). 
We employ the experimental parameters
reported by Gobby {\it et al.} in 
Ref.~\cite{gys}: $Y_0=1.7\times{}10^{-6}$, $e_{d}=0.033$, $\alpha=0.21$ dB/km, and Bob's 
detection efficiency $\eta_{\rm Bob}=0.045$. 
We further assume that  
$q=1$, and $f(E^{c})=f(E^{\bar{c}})=1.22$. These data are used as well for 
simulation purposes in the following sections. 
We 
study two different scenarios: (A) A perfect threshold detector, 
{\it i.e.}, 
$\epsilon=0$ and $\eta_{\rm d}=1$, and (B) 
$\epsilon=3.2\times{}10^{-7}$ and $\eta_{\rm d}=0.12$
\cite{gys}. 
In both cases we find that 
the values of the mean photon number $\mu$ and the transmittance $t$ 
which maximize the secret key rate formula
are quite similar 
and almost constant with the 
distance. In particular, $\mu$ is quite strong (around $200$ in the simulation), while $t$ is quite weak 
(around $10^{-3}$). This result is not surprising.  
When $\mu\gg{}1$ and $t\ll{}1$, Alice's threshold detector produces a click 
most of the times.
Then, in the few occasions  where Alice actually does not see a click in her measurement device, 
she can be quite confident that the signal 
state that goes to Bob is quite weak. Note that in this scenario 
the conditional photon number statistics $q^{\bar{c}}_n$ 
satisfy $q^{\bar{c}}_0\approx{}1$ and $q^{\bar{c}}_{n\geq{}1}\approx{}0$. Similarly to the 
one weak decoy state protocol proposed in Ref.~\cite{estimation}, this fact allows 
Alice and Bob to obtain an accurate estimation of $Y_1$ and $e_1$, which 
results into an 
enhancement of the achievable secret key rate and distance. The cutoff point 
where the secret key rate drops down to zero is $l\approx{}126$ km. 

One can improve the resulting secret key rate further by using 
a passive scheme with more intensity settings.
For instance, Alice may employ
a PNR detector instead of a threshold detector, 
or she could use several threshold detectors in combination with beam splitters. In 
this context, see also Ref.~\cite{mauerer3}.
Figure~\ref{figure3_thermal_combined} illustrates also this last scenario, for 
the case where Alice uses a PNR detector (solid line). As expected, 
it turns out that now the legitimate 
users can estimate the 
actual value of the relevant parameters
$Y_0$, $Y_1$, and $e_1$ with arbitrary precision (see Appendix~\ref{ap_pnr}). 
The cutoff point 
where the secret key rate drops down to zero is $l\approx{}147$ km. 
This result shows that the performance of the passive setup 
represented in Fig.~\ref{figure_general}
with a threshold detector is already close to the best performance 
that can be achieved at all with such an scheme and 
the security analysis provided in Refs.~\cite{gllp,lo_qic}.

\section{Weak coherent light}\label{sec_weak}

Suppose now that the signal states $\rho$ and $\sigma$ which appear in Fig.~\ref{figure_general} are 
two phase randomized WCP emitted by a pulsed laser source. That is,
\begin{eqnarray}\label{rome_cans2}
\rho&=&e^{-\mu_1}\sum_{n=0}^{\infty}\frac{\mu_1^n}{n!}\ket{n}\bra{n}, \nonumber \\
\sigma&=&e^{-\mu_2}\sum_{n=0}^{\infty}\frac{\mu_2^n}{n!}\ket{n}\bra{n},
\end{eqnarray}
with $\mu_1$ and $\mu_2$ denoting, respectively, the mean photon number
of the two signals. 
In this scenario, the joint probability of having $n$ photons in output mode $a$ and $m$ photons in output 
mode $b$ can be written as \cite{mcurty_opt}
\begin{equation}\label{q10}
p_{n,m}=\frac{\upsilon^{n+m}e^{-\upsilon}}{n!m!}\frac{1}{2\pi}\int_0^{2\pi}\gamma^n(1-\gamma)^m d\theta,
\end{equation}
where the parameters $\upsilon$, $\gamma$, and $\xi$, are given by
\begin{eqnarray}
\upsilon&=&\mu_1+\mu_2, \nonumber \\
\gamma&=&\frac{\mu_1{}t+\mu_2{}(1-t)+\xi\cos{\theta}}{\upsilon}, \nonumber \\
\xi&=&2\sqrt{\mu_1\mu_2{}(1-t)t}. 
\end{eqnarray}
This result 
differs from the one expected from the interference of two pure coherent states 
with fixed phase relation, $\ket{\sqrt{\mu_1}e^{i\phi_1}}$ and $\ket{\sqrt{\mu_2}e^{i\phi_2}}$, at a BS
of transmittance $t$. In this last case,
$p_{n,m}$ is just the product of two Poissonian distributions.
Whenever Alice ignores the result of her measurement in mode $b$, then 
the probability of finding  
$n$ photons in mode $a$ can be expressed as 
\begin{equation}\label{help}
p^t_{n}=\sum_{m=0}^{\infty}p_{n,m}=\frac{\upsilon^{n}}{n!}\frac{1}{2\pi}\int_0^{2\pi}\gamma^ne^{-\upsilon{}\gamma} d\theta,
\end{equation}   
which turns out to be a non-Poissonian probability distribution \cite{mcurty_opt}. 
Let us now consider the case where Alice uses a threshold detector to measure output mode $b$. 

\subsection{Threshold detector}\label{coherent_threshold}

\begin{figure}
\centering
\includegraphics[angle=0,scale=0.68]{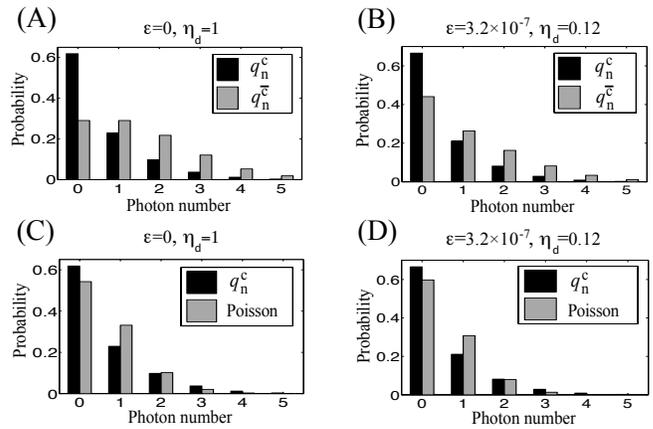}
\caption{
Conditional photon number distribution in mode $a$ (see Fig.~\ref{figure_general}): 
$q^{c}_{n}$ (black) versus $q^{\bar{c}}_{n}$ (grey) when the signal states 
$\rho$ and $\sigma$ are two phase randomized WCP given by Eq.~(\ref{rome_cans2}).
We consider that
$\mu_1=\mu_2=1$ and $t=1/2$, and 
we study two situations: (A) A perfect threshold photon detector, {\it i.e.},  
$\epsilon=0$ and $\eta_{\rm d}=1$ \cite{mcurty_opt}, and (B) $\epsilon=3.2\times{}10^{-7}$ and $\eta_{\rm d}=0.12$. 
These last data correspond to the experiment reported by Gobby {\it et al.} in
Ref.~\cite{gys}. Cases C and D represent $q^{c}_{n}$ (black) versus a Poissonian 
distribution of the same mean photon number for the two scenarios described above
(perfect and imperfect threshold photon detector).} \label{figure2}
\end{figure}

The analysis is completely analogous to the one presented in Sec.~\ref{thermal_threshold}. In particular, 
the joint probability for seeing $n$
photons in mode $a$ and no click in the threshold detector has now the form
\begin{eqnarray}\label{pnc}
p^{\bar{c}}_n&=&(1-\epsilon)\sum_{m=0}^\infty (1-\eta_{\rm d})^mp_{n,m}  \\
&=&(1-\epsilon)\frac{\upsilon^{n}e^{-\eta_{\rm d}\upsilon}}{n!}
\frac{1}{2\pi}\int_0^{2\pi}\gamma^ne^{-(1-\eta_{\rm d})\upsilon\gamma} d\theta. \nonumber
\end{eqnarray}
On the other hand, if the detector produces a click, 
the joint probability of finding $n$ photons in mode $a$ 
is given by Eq.~(\ref{rome_tues}).
Figure~\ref{figure2} (Cases A and B) shows the conditional photon number statistics
of the outcome signal in mode $a$ depending on the result of the detector
(click and no click): 
$q^{c}_{n}=p^{c}_{n}/(1-N_{\rm w})$ and $q^{\bar{c}}_n=p^{\bar{c}}_n/N_{\rm w}$, with
\begin{equation}
N_{\rm w}=\sum_{n=0}^{\infty}p^{\bar{c}}_{n}=(1-\epsilon)
e^{-\eta_{\rm d}[\mu_1{}(1-t)+\mu_2{}t]}I_{0,\eta_{\rm d}\xi}, 
\end{equation}
and where 
$I_{q,z}$ represents the modified Bessel function of the first kind 
\cite{Bessel}. 
This function is defined as \cite{Bessel}
\begin{equation}
I_{q,z}=\frac{1}{2\pi{}i}\oint e^{(z/2)(t+1/t)} t^{-q-1} dt. 
\end{equation}
Figure~\ref{figure2}
includes as well a comparison between $q^{c}_{n}$ and a 
Poissonian distribution of the same mean photon number (Cases C and D). Both distributions, 
$q^{c}_{n}$ and $q^{\bar{c}}_{n}$, are also non-Poissonian. 

\subsection{Lower bound on the secret key rate}

To apply the secret key rate formula given by Eq.~(\ref{ind_kr_new}), with $l\in\{c,\bar{c}\}$, we need 
to estimate the quantities $Y_0$, $Y_1$, and $e_1$. For that, 
we follow the same procedure explained 
in Appendix~\ref{ap_vico1}. 
This method requires that $p^{t}_{n}$ and $p^{\bar{c}}_{n}$ 
satisfy certain conditions that we confirmed numerically. As a result, it turns out that the bounds given by 
Eqs.~(\ref{rome_tues1})-(\ref{eq2}) are also valid in this scenario. 

The only relevant statistics to evaluate Eqs.~(\ref{rome_tues1})-(\ref{eq2}) are 
$p^{t}_n$ and $p^{\bar{c}}_n$, with $n=0,1,2$. These probabilities can be obtained by solving 
Eqs.~(\ref{help})-(\ref{pnc}). They are given in Appendix~\ref{apA}. Note that $p^{c}_n$ 
can be directly calculated from these two statistics by means of Eq.~(\ref{rome_tues}).
After substituting Eqs.~(\ref{yield_new})-(\ref{qber_new})
into the gain and QBER formulas we obtain
\begin{eqnarray}
Q^{\bar{c}}&=&N_{\rm w}-(1-\epsilon)(1-Y_0)e^{(\eta_{\rm d}-\eta_{\rm sys})
\omega-\eta_{\rm d}\upsilon} \nonumber \\
&\times& I_{0,(\eta_{\rm d}-\eta_{\rm sys})\xi}, \nonumber \\
Q^{\bar{c}}E^{\bar{c}}&=&(e_0-e_d)Y_0N_{\rm w}+e_dQ^{\bar{c}}, \nonumber \\
Q^{t}&=&1-(1-Y_0)e^{-\eta_{\rm sys}\omega}I_{0,\eta_{\rm sys}\xi}, \nonumber \\
Q^{t}E^{t}&=&(e_0-e_d)Y_0+e_dQ^{t},
\end{eqnarray}
with the parameter $\omega$ given by
\begin{equation}
\omega=\mu_1{}t+\mu_2(1-t).
\end{equation}

\begin{figure}
\centering
\includegraphics[angle=0,scale=0.35]{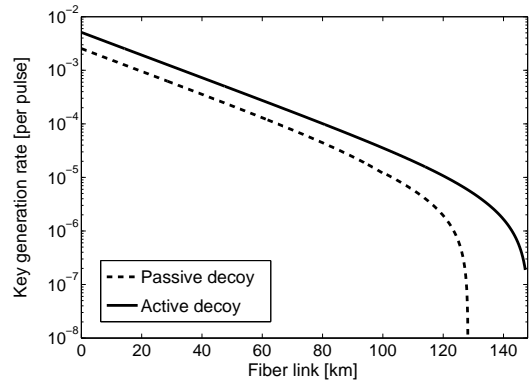}
\caption{
Lower bound on the secret key rate $R$ given by Eq.~(\ref{key_rate}) 
in logarithmic scale 
for the passive decoy state setup illustrated in Fig.~\ref{figure_general} 
with two intensity settings.  
The signal states $\rho$ and $\sigma$ are two phase randomized WCP given by 
Eq.~(\ref{rome_cans2}). 
The transmittance of the BS is $t=1/2$.
We consider two possible scenarios: (A) $\epsilon=0$ and $\eta_{\rm d}=1$ \cite{mcurty_opt} ({\it i.e.}, a perfect 
threshold photon detector), 
and (B) $\epsilon=3.2\times{}10^{-7}$ and $\eta_{\rm d}=0.12$ \cite{gys}. Both cases provide approximately 
the same final key rate and they cannot be distinguished with the resolution of this figure (dashed line). The solid line
represents a lower bound on $R$ for an active asymptotic decoy state system with infinite 
decoy settings
\cite{decoy2}.
This last result coincides approximately with the case where Alice employs a PNR detector (see
Appendix~\ref{ap_pnr2}), 
and the secret key rate is both scenarios cannot be distinguished with the resolution of this figure.} 
\label{figure3}
\end{figure}

The resulting lower bound on the secret key rate is illustrated in Fig.~\ref{figure3}. 
We assume that  $t=1/2$, {\it i.e.}, 
we consider a simple $50:50$ BS.  
Again, 
we study two different situations: (A)  
$\epsilon=0$ and $\eta_{\rm d}=1$ \cite{mcurty_opt}, and (B) 
$\epsilon=3.2\times{}10^{-7}$ and $\eta_{\rm d}=0.12$
\cite{gys}. In both cases 
the optimal values of the intensities $\mu_1$ and $\mu_2$ are almost constant with the 
distance. One of them is quite weak (around $10^{-4}$), while the other one is around 
$0.5$. 
The reason for this result can be understood as follows. When the intensity of one 
of the signals is really weak, 
the output photon number distributions in mode $a$ are always close 
to a Poissonian distribution (for click and no click events). This distribution is narrower than
the one arising when both $\mu_1$ and $\mu_2$ are of the same order of magnitude.
In this case, a better estimation of $Y_1$ and $e_1$ can be derived, and this fact 
translates into a higher secret key rate. It must be emphasized, however, that 
from an experimental point of view this solution might not be optimal. Specially, 
since in this scenario the two output distributions $p_n^c$ and $p_n^{\bar{c}}$
might be too close to each other for being distinguished in practice. 
This effect could be specially relevant when one considers statistical fluctuations due to finite 
data size (see Sec.~\ref{fluc}). For instance, 
small fluctuations in a practical system could overwhelm the tiny difference between 
the decoy state and the signal state in this case.
Figure~\ref{figure3} includes as well the secret key rate of an 
active asymptotic decoy state QKD system with infinite decoy settings \cite{decoy2}. 
The cutoff points where the secret key rate drops down to zero are $l\approx{}128$ km (passive setup
with two intensity settings)
and $l\approx{}147$ km (active 
asymptotic setup). 
From these results we see that the performance of 
the passive scheme with a threshold detector is comparable to the active one, thus showing the practical 
interest of the passive setup.  

Like in Sec.~\ref{sec_thermal}, one can improve the performance of the passive scheme further 
by using more intensity settings. The case where Alice uses a 
PNR detector is analyzed in Appendix~\ref{ap_pnr2}. The result is 
also shown in Fig.~\ref{figure3}. It reproduces approximately the behavior of the asymptotic
active setup and the secret key rate is both scenarios cannot be distinguished with the resolution of this figure 
(solid line). This result is not surprising, since in 
both situations (passive and active) we apply Eq.~(\ref{ind_kr_new}) with the actual values of the 
parameters $Y_0$, $Y_1$, and $e_1$. The only difference 
between these two setups 
arises from the photon 
number distribution of the signal states that go to Bob. In particular, 
while in the passive scheme the relevant statistics are given by Eq.~(\ref{rome_bar}), 
in the
active setup these statistics have the form given by Eq.~(\ref{nap1}).

\subsection{Alternative implementation scheme}\label{sec_alternative}

The passive setup illustrated in Fig.~\ref{figure_general} requires that 
Alice employs two independent sources of signal states. This fact might become 
specially relevant when she uses phase randomized
WCP, since in this situation none of the signal states entering the BS can be the vacuum state. Otherwise, 
the photon number distributions of the output signals in mode $a$ and mode $b$
would be statistically independent. 

\begin{figure}
\begin{center}
\includegraphics[angle=0,scale=0.62]{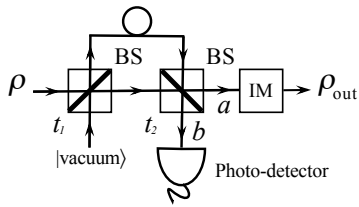}
\end{center}
\caption{
Alternative implementation scheme with only one pulsed laser source. 
The delay introduced by one arm of the interferometer is equal to the time difference 
between two pulses. The intensity modulator (IM) blocks either all the even or all the odd 
optical pulses in mode $a$.
\label{figure1_new}}
\end{figure}

Alternatively to the passive scheme shown in Fig.~\ref{figure_general}, Alice could as well
employ, 
for instance, the scheme
illustrated in Fig.~\ref{figure1_new}.
This setup has
only one laser diode, but follows a similar spirit like the original scheme in Fig.~\ref{figure_general}, where a 
photo-detector is used to measure the output signals in mode $b$. It includes, however, 
an intensity modulator (IM) to block either all the even or all the odd pulses in mode $a$. 
This requires, therefore, an active control of the functioning of the IM, but note that 
no random number generator is needed here. 
The main reason for blocking half of the pulses in mode $a$ is to suppress possible correlations between 
them. That is, the action of the IM  guarantees that the signal states that go to Bob are
tensor product of mixtures of Fock states. Then, one can directly apply the security 
analysis provided in Refs.~\cite{decoy2,gllp,lo_qic}.
Thanks to the one-pulse delay introduced by one arm of the interferometer, together with a proper selection of 
the transmittance $t_1$, it can be shown that both setups in Fig.~\ref{figure_general} and
Fig.~\ref{figure1_new} are completely equivalent, except 
from the resulting secret key rate. More precisely, the secret key rate in the active scheme is 
half the one of the passive setup, since half of
the pulses are now discarded. 

\section{Strong coherent light}\label{sec_strong}

\begin{figure}
\begin{center}
\includegraphics[angle=0,scale=0.62]{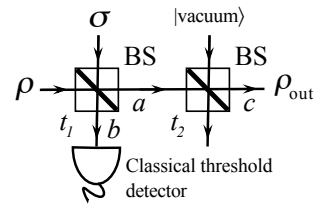}
\end{center}
\caption{
Basic setup of a passive decoy state QKD scheme with strong coherent light. The mean photon 
number of the signal states $\rho$ and $\sigma$ is now quite high; for instance, around $\approx{}10^8$ photons.
$t_1$ and $t_2$ represent the transmittances of the two BS, and $a$, $b$, and $c$ 
denote output modes.
\label{figure1_scp}}
\end{figure}

Let us now consider the passive decoy state setup illustrated in Fig.~\ref{figure1_scp}.
This scheme presents two 
main differences with respect to the passive system analyzed in Sec.~\ref{sec_weak}. In particular,  
the mean photon number (intensity) of the signal states $\rho$ and $\sigma$ is now 
very high; for instance, $\approx{}10^8$ photons. This fact allows Alice to use a simple classical 
photo-detector 
to measure the pulses in mode $b$, which makes this scheme specially suited for experimental 
implementations.
Moreover, it has an additional BS of transmittance $t_2$
to attenuate the signal states in mode $a$ and bring them to the QKD regimen. 

Due to the high intensity of the input signal states $\rho$ and $\sigma$, we can describe the action of 
the first BS
in Fig.~\ref{figure1_scp} by means of a classical model. Specifically, let
$I_1$ ($I_2$) represent the intensity of the input
states $\rho$ ($\sigma$), and let $I_a(\theta)$ [$I_b(\theta)$] be
the intensity of the output pulses in mode $a$ ($b$). Here the angle $\theta$ 
is just a function of the relative phase between the two input states. It
is given by 
\begin{equation}
\theta=\phi_1-\phi_2+\pi/2, 
\end{equation}
where $\phi_1$ ($\phi_2$) denotes the phase of the signal $\rho$ ($\sigma$). Like in 
Sec.~\ref{sec_weak}, we
assume that these phases are uniformly distributed between $0$ and $2\pi$ for each pair of 
input states. This can be achieved, for instance, if Alice uses two pulsed laser sources
to prepare the signals $\rho$ and $\sigma$. 
With this notation, we have that
$I_a(\theta)$ and $I_b(\theta)$ can be expressed as
\begin{eqnarray}
I_a(\theta)&=&t_1I_1+r_1I_2+2\sqrt{t_1r_1I_1I_2}\cos{\theta}, \nonumber \\
I_b(\theta)&=&r_1I_1+t_1I_2-2\sqrt{t_1r_1I_1I_2}\cos{\theta},
\end{eqnarray}
where $t_1$ denotes the transmittance of the BS, and
$r_1=1-t_1$.

\subsection{Classical threshold detector}

\begin{figure}
\centering
\includegraphics[angle=0,scale=0.32]{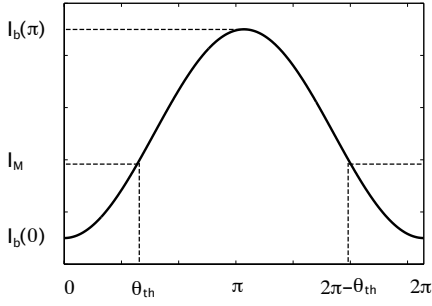}
\caption{Graphical representation of the intensity $I_b(\theta)$ in mode $b$ (see Fig.~\ref{figure1_scp}) versus 
the angle $\theta$. $I_M$ represents the threshold value of the 
classical threshold detector, and $\theta_{\rm th}$
is its associated threshold angle.} \label{figure_Ib}
\end{figure}

For simplicity, we shall consider that Alice uses a {\it perfect} 
classical threshold detector to measure the pulses in 
mode $b$. For each incoming signal, this device tells her whether its intensity is below or above a certain 
threshold value $I_M$ that satisfies $I_b(\pi)>I_M>I_b(0)$. That is, the value of 
$I_M$ is between the minimal and maximal 
possible values of the intensity of the pulses in mode $b$. Note, however, that the analysis presented 
in this section 
can be straightforwardly adapted to cover also the case of an imperfect classical threshold detector, or 
a classical photo-detector with several threshold settings. 
Figure~\ref{figure_Ib} shows a graphical representation of $I_b(\theta)$ versus the angle
$\theta$, together with the threshold value $I_M$. 
The angle $\theta_{\rm th}$ which satisfies
$I_b(\theta_{\rm th})=I_M$ is given by  
\begin{equation}
\theta_{\rm th}=\arccos{\Bigg(\frac{r_1I_1+t_1I_2-I_M}{2\sqrt{t_1r_1I_1I_2}}\Bigg)}.
\end{equation}

\begin{figure}
\begin{center}
\includegraphics[angle=0,scale=0.72]{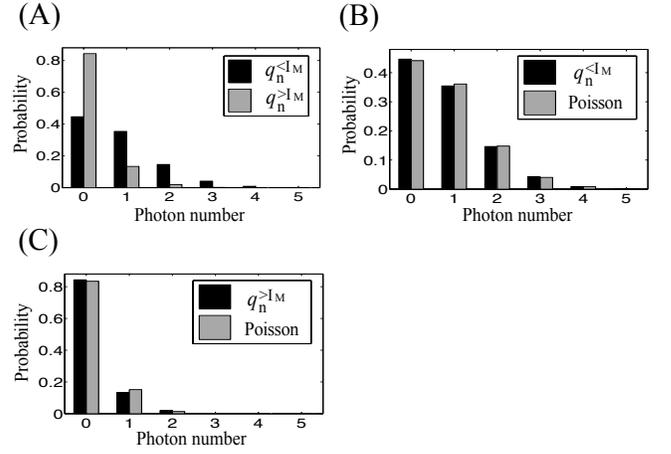}
\end{center}
\caption{
(A) Conditional photon number distribution in mode $c$ (see Fig.~\ref{figure1_scp}): 
$q_n^{<I_M}$ (black) and $q_n^{>I_M}$ (grey) for the case $I_1=I_2=I_M=10^8$, $t_1=1/2$, and
$t_2=0.5\times{}10^{-8}$.
Cases B and C represent, respectively, $q_n^{<I_M}$ and $q_n^{>I_M}$ (black) versus a Poissonian 
distribution of the same mean photon number (grey).
\label{figure2_scp}}
\end{figure}

Whenever the classical threshold detector provides Alice with an intensity value below $I_M$, it turns 
out that the unnormalized 
signal states in mode $c$ can be expressed as 
\begin{eqnarray}
\rho_{\rm out}^{<I_M}&=&\frac{1}{2\pi}\sum_{n=0}^\infty\Bigg\{\int_0^{\theta_{\rm th}} 
\frac{e^{-I_a(\theta)t_2}[I_a(\theta)t_2]^n}{n!}\ket{n}\bra{n}d\theta \nonumber \\
&+&\int_{2\pi-\theta_{\rm th}}^{2\pi} 
\frac{e^{-I_a(\theta)t_2}[I_a(\theta)t_2]^n}{n!}\ket{n}\bra{n}d\theta\Bigg\} \nonumber \\
&=&\frac{1}{\pi}\sum_{n=0}^\infty\int_0^{\theta_{\rm th}} 
\frac{e^{-I_a(\theta)t_2}[I_a(\theta)t_2]^n}{n!}\ket{n}\bra{n}d\theta.
\end{eqnarray}
This means, in particular, that the joint probability of finding $n$ photons in mode $c$ and 
an intensity value below $I_M$ in mode $b$  is given by
\begin{equation}\label{rome_wed1}
p_n^{<I_M}=\frac{t_2^n}{n!\pi}\int_{0}^{\theta_{\rm th}} I_a(\theta)^n e^{-I_a(\theta)t_2} d\theta.
\end{equation}
Similarly, we find that $p_n^{>I_M}$ can be written as
\begin{equation}\label{rome_wed2}
p_n^{>I_M}=\frac{t_2^n}{n!\pi}\int_{\theta_{\rm th}}^{\pi} I_a(\theta)^n e^{-I_a(\theta)t_2} d\theta.
\end{equation}
Figure~\ref{figure2_scp} (Case A) shows the conditional photon number statistics of the 
outcome signal 
in mode $c$ depending on the result of the classical threshold detector (below or above $I_M$): 
$q^{<I_M}_{n}=p^{<I_M}_{n}/N_{\rm s}$ and $q^{>I_M}_n=p^{>I_M}_n/(1-N_{\rm s})$, with
\begin{equation}
N_{\rm s}=\sum_{n=0}^{\infty}p^{<I_M}_{n}=\frac{\theta_{\rm th}}{\pi}.
\end{equation}
This figure
includes as well a comparison between $q^{<I_M}_{n}$ (Case B) and $q^{>I_M}_{n}$ (Case C) and a 
Poissonian distribution of the same mean photon number. It turns out that both distributions, 
$q^{<I_M}_{n}$ and $q^{>I_M}_{n}$, approach a Poissonian distribution when $t_2$ is 
sufficiently small.

\subsection{Lower bound on the secret key rate}

Again, to apply the secret key rate formula given by Eq.~(\ref{ind_kr_new}), 
with $l\in\{<I_M,>I_M\}$, we need 
to estimate the quantities
$Y_0$, $Y_1$, and $e_1$.
Once more, we follow the procedure explained 
in Appendix~\ref{ap_vico1}. 
We confirmed numerically that the probabilities $p_n^{<I_M}$ and $p_n^{>I_M}$ 
satisfy the conditions required to use this technique.
As a result, it turns out that the bounds given by 
Eqs.~(\ref{rome_tues1})-(\ref{eq2}) are also valid in this scenario. 

For simplicity, we impose $I_1=I_2=I_M\equiv{}I$. This means that $\theta_{\rm th}=\pi/2$.
The relevant statistics 
$p_n^{<I_M}$ and $p_n^{>I_M}$, with $n=0,1,2$, are calculated in Appendix~\ref{apB}. 
After substituting Eqs.~(\ref{yield_new})-(\ref{qber_new})
into the gain and QBER formulas we obtain
\begin{eqnarray}
Q^{<I_M}&=&N_{\rm s}-\frac{(1-Y_0)e^{-\eta_{\rm sys}\kappa}}{2}(I_{0,\eta_{\rm sys}\zeta}-L_{0,\eta_{\rm sys}\zeta}), \nonumber \\
Q^{<I_M}E^{<I_M}&=&(e_0-e_d)Y_0N_{\rm s}+e_dQ^{<I_M}, \nonumber \\
Q^{>I_M}&=&(1-N_{\rm s})-\frac{(1-Y_0)e^{-\eta_{\rm sys}\kappa}}{2} \nonumber \\
&\times&(I_{0,\eta_{\rm sys}\zeta}+L_{0,\eta_{\rm sys}\zeta}), \nonumber \\
Q^{>I_M}E^{>I_M}&=&(e_0-e_d)Y_0(1-N_{\rm s})+e_dQ^{>I_M},
\end{eqnarray}
where the parameter $\kappa$ is given by 
\begin{equation}
\kappa=It_2, 
\end{equation}
and $L_{q,z}$ represents the modified Struve function \cite{Bessel2} defined by Eq.~(\ref{struve}).

\begin{figure}
\begin{center}
\includegraphics[angle=0,scale=0.35]{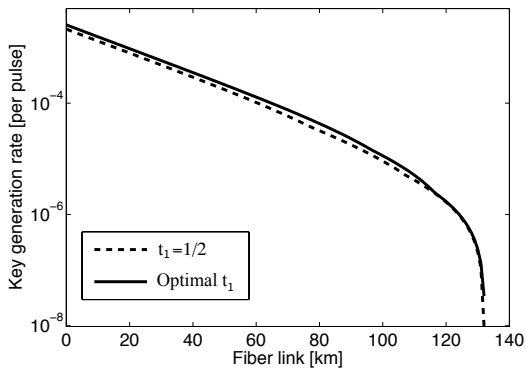}
\end{center}
\caption{Lower bound on the secret key rate $R$ given by Eq.~(\ref{key_rate}) 
in logarithmic scale 
for the passive decoy state setup illustrated in Fig.~\ref{figure1_scp}
with two intensity settings.  
We consider two possible scenarios: (A)
We impose $t_1=1/2$, {\it i.e.}, 
we consider a simple $50:50$ BS, and we optimize the parameter $\kappa$ (dashed line), 
and (B) we optimize 
both parameters, $t_1$ and $\kappa$ (solid line). 
\label{figure3_scp}}
\end{figure}

The resulting lower bound on the secret key rate is illustrated in Fig.~\ref{figure3_scp}. 
We study two different situations: (A) We impose $t_1=1/2$, {\it i.e.}, 
we consider a simple $50:50$ BS, and we optimize the parameter $\kappa$, and (B) we optimize 
both quantities, $t_1$ and $\kappa$. 
In both scenarios
the optimal values of the parameters are almost constant with the 
distance. In the first case $\kappa$ is around $0.2$, while in the second case 
we obtain that  $t_1$ and $\kappa$ are, respectively, around $0.06$ and $0.25$.
The cutoff point where the secret key rate drops down to zero is $l\approx{}132$ km both in case A and B.
These results seem to indicate that this passive scheme can offer a better performance 
than the passive setups analyzed in Sec.~\ref{sec_thermal} and in Sec.~\ref{sec_weak} with a threshold
photon detector. This fact arises mainly from the probability distributions $p_n^{<I_M}$ and $p_n^{>I_M}$, which,
in this scenario, approach a Poissonian distribution when $t_2$ is 
sufficiently small. Again, one can improve the performance of this system even further just by using 
more threshold settings in the classical threshold detector. 
Moreover, from an experimental point of view, this configutation
might be more feasible than using PNR detectors. 

\begin{figure}
\begin{center}
\includegraphics[angle=0,scale=0.62]{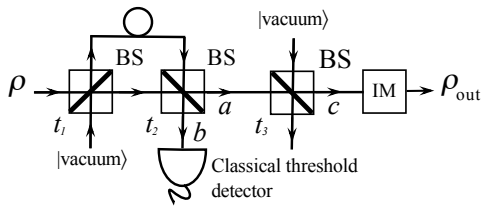}
\end{center}
\caption{
Alternative implementation scheme with only one pulsed laser source. 
The delay introduced by one arm of the interferometer is equal to the time difference 
between two pulses. The intensity modulator (IM) blocks either the even or the odd 
optical pulses in mode $c$. 
\label{figure1_new_scp}}
\end{figure}

To conclude this section, let us mention that, like in Sec.~\ref{sec_alternative}, Alice could as well
employ, 
for instance, the alternative active scheme
illustrated in Fig.~\ref{figure1_new_scp}.
This setup has
only one pulsed laser source, but includes 
an intensity modulator (IM) to block either all the even or all the odd pulses in mode $c$.
The argumentation here goes exactly the same like in Sec.~\ref{sec_alternative} and we omit it
for simplicity. The resulting secret key rate in the active scheme 
is half the one of the passive setup. 

\section{Statistical Fluctuations}\label{fluc}

In this section, we discuss briefly the effect that finite data size in real life experiments might have on the final 
secret key rate. For that, we follow the statistical fluctuation analysis presented in Ref.~\cite{estimation}. This 
procedure is based on standard error analysis. That is, we shall assume that all the variables which are measured in 
the experiment each fluctuates around its asymptotic value.

Our main objective here is to obtain a lower bound on the secret 
key rate formula given by Eq.~(\ref{ind_kr_new}) under statistical fluctuations. 
For that, we realize the following four assumptions: 
\begin{enumerate}
\item Alice and Bob know the photon number statistics of the 
source well and we do not consider their fluctuations directly. 
Intuitively speaking, these fluctuations are included
in the parameters
measuring the gains and QBERs.
\item Alice and Bob use a real upper bound on the single photon error rate $e_1$, 
thus no fluctuations have to be considered for this parameter. In particular, we use the fact that the number of errors 
within the single photon states cannot be greater than the total number of errors.
\item Alice and Bob use a
standard error analysis procedure to deal with the fluctuations of the variables which are measured. 
\item The error rate of background does not fluctuate, {\it i.e.}, $e_0=1/2$.
\end{enumerate}

To illustrate our results, we focus on the passive decoy state setup introduced in Sec.~\ref{sec_weak}. Note, 
however, that a similar 
analysis can also be applied to the other passive schemes presented in this paper. 

\subsection{Active decoy state QKD}
In order to make a fair comparison between the active and the passive decoy state QKD setups with 
two intensity settings, 
from now on we shall consider an active scheme with only one decoy state \cite{estimation}. 
In this last case, the quantities
$Y_1$ and $e_1$ can be bounded as
\begin{eqnarray} \label{Fluc:Ye1act}
Y_1&\geq{}&Y_1^L= \frac{\mu^2Q_\nu e^\nu-\nu^2Q_\mu e^\mu-(\mu^2-\nu^2)Y_0}{\mu\nu(\mu-\nu)}, \nonumber \\
e_1&\leq{}&e_1^U = \frac{E_\mu Q_\mu e^\mu-e_0Y_0}{Y_1^l \mu}, 
\end{eqnarray}
where $\mu$ ($\nu$) denotes the mean photon number of a signal (decoy) state,
$Q_\mu$ ($Q_\nu$) and $E_\mu$ ($E_\nu$) represent, respectively, its associated gain and 
QBER, and $Y_0$ is a free parameter. 
Using the channel model described in Sec.~\ref{sec_model}, we find that 
these parameters can be written as 
\begin{equation} \label{Fluc:QEmuact}
\begin{aligned}
Q_\mu &= Y_0+1-e^{-\mu\eta_{\rm sys}}, \\
E_\mu Q_\mu &= e_0Y_0+e_d(1-e^{-\mu\eta_{\rm sys}}), \\
Q_\nu &= Y_0+1-e^{-\nu\eta_{\rm sys}}, \\
E_\nu Q_\nu &= e_0Y_0+e_d(1-e^{-\nu\eta_{\rm sys}}). \\
\end{aligned}
\end{equation}
If we now apply a 
standard error analysis to these quantities we obtain that their deviations from the theoretical values
are given by
\begin{equation} \label{Fluc:DeltaQmunuact}
\begin{aligned}
\Delta_{Q\mu} &= u_\alpha \sqrt{Q_\mu/N_\mu}, \\
\Delta_{Q\nu} &= u_\alpha \sqrt{Q_\nu/N_\nu}, \\
\Delta_{Q\mu E\mu} &= u_\alpha \sqrt{2E_\mu Q_\mu/N_\mu}, \\
\Delta_{Q\nu E\nu} &= u_\alpha \sqrt{2E_\nu Q_\nu/N_\nu}, \\
\end{aligned}
\end{equation}  
where $N_\mu$ ($N_\nu$) denotes 
the number of signal (weak decoy) pulses sent by Alice, and $u_\alpha$ represents the number 
standard deviations from the central values. That is, the total number of pulses emitted by the source
is just given by 
$N=N_\mu+N_\nu$.  
Roughly speaking, this means, for instance, that the gain of the signal states lies in the interval 
$Q_\mu\pm\Delta_{Q\mu}$ except
with small probability, and similarly for the other quantities 
defined in Eq.~(\ref{Fluc:QEmuact}).
For example, if we select $u_\alpha=10$, then the corresponding confidence 
interval is $1-1.5\times10^{-23}$, which 
we use later on for simulation purposes.
For simplicity, here 
we have assumed that Alice and Bob use the standard BB84 protocol, {\it i.e.},
they keep only half of their raw bits (due to the basis sift). 
This is the reason for the factor $2$ which appears in the last two expressions of 
Eq.~(\ref{Fluc:DeltaQmunuact}). In this context, see also 
Ref.~\cite{hay} for a discussion on the optimal value of the parameter $q$. 

\subsection{The background $Y_0$} \label{Sub:Y0}

The bounds given by Eq.~(\ref{Fluc:Ye1act}) depend on the unknown parameter $Y_0$. When 
a vacuum decoy state is applied, the value of $Y_0$ can be estimated. Alternatively, 
one can also derive a lower bound on $Y_1$ and an upper bound on $e_1$ which do not 
depend on $Y_0$. Specifically, 
from  
Eqs.~(\ref{gain_rome})-(\ref{rome3}) we obtain that
\begin{eqnarray} \label{Fluc:Y1L1}
(1-2e_1)Y_1 \geq A &=& \frac{\mu}{\nu(\mu-\nu)}Q_\nu (1-2E_\nu) e^\nu \nonumber \\
&-&\frac{\nu}{\mu(\mu-\nu)}Q_\mu (1-2E_\mu) e^\mu.
\end{eqnarray}
The gains $Q_\mu$ and $Q_\nu$, together with the QBERs $E_\mu$ and $E_\nu$, are directly measured 
in the experiment, and their statistical fluctuations are given by Eq.~\eqref{Fluc:DeltaQmunuact}. On the other hand, 
we have that
\begin{equation} \label{Fluc:e1U}
e_1 \leq \frac{B}{Y_1^L},
\end{equation}
with the parameter $B$ given by
\begin{equation}
B = \min\Bigg\{\frac{E_\nu Q_\nu e^\nu}{\nu}, \frac{E_\mu Q_\mu e^\mu-E_\nu Q_\nu e^\nu}{\mu-\nu}\Bigg\}.
\end{equation}
Combining Eqs.~\eqref{Fluc:Y1L1}-\eqref{Fluc:e1U} we find 
\begin{equation}\label{Fluc:KeyPri}
Y_1[1-H(e_1)] \geq \frac{A}{1-2e_1}\Bigg[1-H\Bigg(\frac{B(1-2e_1)}{A}\Bigg)\Bigg]. 
\end{equation}
The quantities $A$ and $B$ can be obtained directly from the variables measured in the 
experiment. Moreover, if one considers the 
secret key rate 
formula given by Eq.~(\ref{ind_kr_new})
as a function of the free parameter $e_1$, then one should select an upper bound on 
$e_1$, which gives a value (may not be a bound) for $Y_1$ as
\begin{eqnarray} \label{Fluc:Y1e1new}
Y_1^t &=& A+2B, \nonumber \\
e_1^U &=& \frac{B}{A+2B},
\end{eqnarray}
where the equation for $e_1^U$ comes from solving the two inequalities given by  
Eqs.~\eqref{Fluc:Y1L1}-\eqref{Fluc:e1U}.

Again, using a standard error analysis procedure, we find that the deviations of the 
parameters $A$ and $B$ from their theoretical values can be written as
\begin{eqnarray}\label{m_i}
\Delta_{A} &=& \Big[(c_1\Delta_{Q\nu})^2+4(c_1\Delta_{E\nu Q\nu})^2 +(c_2\Delta_{Q\mu})^2 \nonumber \\
&&+
4(c_2\Delta_{E\mu Q\mu})^2\Big]^{\frac{1}{2}}, \nonumber \\
\Delta_{B} &=& \min\Bigg\{\frac{e^\mu\Delta_{E\mu Q\mu}}{\mu},  \frac{e^\nu\Delta_{E\nu Q\nu}}{\nu},\nonumber \\
&& \frac{\sqrt{(e^\mu\Delta_{E\mu Q\mu})^2+(e^\nu\Delta_{E\nu Q\nu})^2}}{\mu-\nu}
\Bigg\},
\end{eqnarray}
where the coefficients $c_1$ and $c_2$ have the form
\begin{eqnarray}
c_1 &=& \frac{\mu}{\nu(\mu-\nu)}e^\nu, \nonumber \\
c_2 &=& \frac{\nu}{\mu(\mu-\nu)}e^\mu,
\end{eqnarray}
and the deviations of the gains and the QBERs are given by Eq.~\eqref{Fluc:DeltaQmunuact}.

For simplicity, we assume now that $A$ and $B$ are statistically independent. Thus, the statistical 
deviation of the crucial term $Y_1[1-H_2(e_1)]$ in 
the secret key formula can be written as 
\begin{eqnarray} \label{Fluc:KeyPrifluc}
\Delta_{Y_1[1-H_2(e_1)]} &=& \Bigg\{\Bigg[\Delta_{A}\log_2\Bigg(\frac{2A+2B}{A+2B}\Bigg)\Bigg]^2 \\
&+&\Bigg[\Delta_{B}\log_2\Bigg(\frac{4B(A+B)}{(A+2B)^2}\Bigg)\Bigg]^2\Bigg\}^{\frac{1}{2}}.  \nonumber 
\end{eqnarray}

\begin{figure}
\begin{center}
\includegraphics[angle=0,scale=0.35]{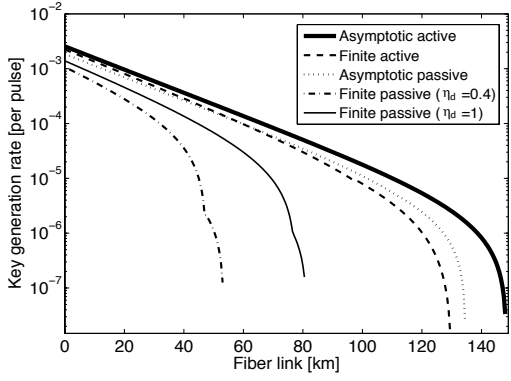}
\end{center}
\caption{Lower bound on the secret key rate $R$ given by Eq.~(\ref{key_rate})  
in logarithmic scale. We consider four possible scenarios: (A) An active decoy state setup 
(with only one decoy state) with
statistical fluctuations (dashed line) \cite{estimation}, 
(B) An active decoy state setup 
(with only one decoy state)
without considering statistical fluctuations (thick solid line) \cite{estimation},
(C) The passive decoy state scheme with WCP introduced in Sec.~\ref{sec_weak} now considering
statistical fluctuations. In this last case, moreover, 
we study two possible situations depending on the value of $\eta_{\rm d}$: 
(C1) $\eta_{\rm d}=1$ (thin solid line), and (C2)
$\eta_{\rm d}=0.4$ (dash-dotted line). (D) The passive decoy state scheme with WCP introduced 
in Sec.~\ref{sec_weak} with $\eta_{\rm d}=1$ and without considering 
statistical fluctuations (dotted line). In all passive setups 
the transmittance of the BS is $t=1/2$ and we use $\epsilon=0$. 
Furthermore, we pick the data size (total number of pulses emitted by Alice) to 
be $N=6\times10^9$. The confidence 
interval for statistical fluctuations is ten standard deviations ({\it i.e.}, $1-1.5\times10^{-23}$).
\label{vgo1}}
\end{figure}

From Eqs.~\eqref{Fluc:DeltaQmunuact}, \eqref{m_i} and \eqref{Fluc:KeyPrifluc} one can directly calculate the 
final secret key rate with statistical fluctuations for an active decoy state setup with only one decoy state 
\cite{estimation}. The result 
is illustrated in Fig.~\ref{vgo1} (dashed line). Here we use again the experimental data
reported by Gobby {\it et al.} in Ref.~\cite{gys}. Moreover,
we pick the data size (total number of pulses emitted by Alice) to 
be $N=6\times10^9$.
We calculate the 
optimal values of 
$\mu$ and $\nu$ for each fiber length numerically. It turns out that 
both parameters are almost constant with the distance. One of them is weak 
(it varies between $0.03$ and $0.06$), while the other is around $0.48$. 
This figure 
includes as well the resulting secret key rate for the same setup without considering 
statistical fluctuations (thick solid line). 
The cutoff points where the secret key rate drops down to zero are $l\approx{}129.5$ km (active setup
with statistical fluctuations)
and $l\approx{}147$ km (active setup
without considering statistical fluctuations). 
From these results we see that the performance of 
this active scheme is quite robust against statistical fluctuations.

\subsection{Passive decoy state QKD}

The analysis is completely analogous to the previous section. Specifically, we find 
that the parameters $A$ and $B$ are now given by
\begin{eqnarray}
A &=& \frac{p^{\bar{c}}_2Q^t(1-2E^t)-p^t_2Q^{\bar{c}}(1-2E^{\bar{c}})} {p^{\bar{c}}_2p^t_1-p^t_2p^{\bar{c}}_1}, \nonumber \\
B &=& \min\Bigg\{\frac{E^{\bar{c}}Q^{\bar{c}}}{p^{\bar{c}}_1}, \frac{p_0^{\bar{c}}E^tQ^t-p_0^tE^{\bar{c}}Q^{\bar{c}}}
{p_0^{\bar{c}}p^t_1-p_0^tp^{\bar{c}}_1}\Bigg\},
\end{eqnarray}
while Eq.~(\ref{Fluc:Y1e1new}) is still valid in this scenario. The deviations of $A$ and $B$ 
have the form
\begin{eqnarray}
\Delta_{A} &=& \frac{1}{p^{\bar{c}}_2p^t_1-p^t_2p^{\bar{c}}_1}
\Big[(p^{\bar{c}}_2\Delta_{Q^t})^2+4(p^{\bar{c}}_2\Delta_{E^tQ^t})^2 \nonumber \\
&&+(p^t_2\Delta_{Q^{\bar{c}}})^2+4(p^t_2\Delta_{E^{\bar{c}}Q^{\bar{c}}}\Big]^{\frac{1}{2}}, \nonumber \\ 
\Delta_{B} &=& \min\Bigg\{\frac{\Delta_{E^tQ^t}}{p^t_1}, \frac{\Delta_{E^{\bar{c}}Q^{\bar{c}}}}{p^{\bar{c}}_1},\nonumber \\
&&\frac{\sqrt{(p_0^{\bar{c}}\Delta_{E^tQ^t})^2+(p_0^t\Delta_{E^{\bar{c}}Q^{\bar{c}}})^2}}
{p_0^{\bar{c}}p^t_1-p_0^tp^{\bar{c}}_1}\Bigg\}.
\end{eqnarray}
On the other hand, the deviations of the gains and the QBERs can now be written as
\begin{equation} \label{Fluc:ABfluc}
\begin{aligned}
\Delta_{Q^t} &= u_\alpha\sqrt{Q^t/N}, \\
\Delta_{Q^{\bar{c}}} &= u_\alpha\sqrt{Q^{\bar{c}}/N^{\bar{c}}}, \\
\Delta_{E^tQ^t} &= u_\alpha\sqrt{2E^t Q^t/N}, \\
\Delta_{E^{\bar{c}}Q^{\bar{c}}} &= u_\alpha\sqrt{2E^{\bar{c}} Q^{\bar{c}}/N^{\bar{c}}}, \\
\end{aligned}
\end{equation}
where $N^{\bar{c}}$ denotes the number of pulses where Alice obtained no click 
in her threshold detector, and $N$ is the total number of pulses emitted by the source. 
The deviation of the term $Y_1[1-H_2(e_1)]$ is again given by Eq.~(\ref{Fluc:KeyPrifluc}).

The secret key rate for the 
passive decoy state scheme with WCP 
introduced in Sec.~\ref{sec_weak} with two intensity settings
and considering statistical fluctuations is illustrated in Fig.~\ref{vgo1}.
We assume that  $t=1/2$, {\it i.e.}, 
we consider a simple $50:50$ BS, and $\epsilon=0$.  
The data size is equal to the one of the previous section, {\it i.e.}, 
$N=6\times10^9$.
We study two different situations depending on the
efficiency of Alice's threshold detector:
$\eta_{\rm d}=1$ (thin solid line), and 
$\eta_{\rm d}=0.4$ (dash-dotted line). 
In both cases the optimal values of the intensities 
$\mu_1$ and $\mu_2$ are almost constant with the distance. One 
of them is weak (it varies between $0.1$ and $0.17$), while the 
other is around $0.5$. 
Figure~\ref{vgo1} 
includes as well the resulting secret key rate for the same setup with $\eta_{\rm d}=1$ 
and without considering statistical fluctuations (dotted line).
The cutoff points where the secret key rate drops down to zero are $l\approx{}53$ km (passive setup
with statistical fluctuations and $\eta_{\rm d}=0.4$), 
$l\approx{}80$ km (passive setup
with statistical fluctuations and $\eta_{\rm d}=1$),
and $l\approx{}128$ km (passive setup
without considering statistical fluctuations, see Sec.~\ref{sec_weak}). From these
results we see that the performance of the passive schemes 
introduced in Sec.~\ref{sec_weak}
(with statistical fluctuations) depends on the actual 
value of the efficiency $\eta_{\rm d}$. In particular, 
when Alice's detector efficiency
is low, the photon number statistics of the signal states that go to Bob
(conditioned on Alice's detection)
become close to each other. This effect becomes specially relevant when 
one considers statistical fluctuations due to finite data size. In this last case, 
small fluctuations can easily cover the difference between the signal states 
associated, respectively, to click and no click events on Alice's threshold
detector. As a result, the achievable secret key rate and distance decrease.

\section{Conclusion}\label{conc}

In this paper we have extended the results presented in Ref.~\cite{mcurty_opt}, 
now showing specifically the 
analysis for other practical scenarios with different light sources and photo-detectors. 
In particular, we have considered sources 
emitting thermal states and phase randomized 
WCP in combination with threshold detectors and photon number 
resolving (PNR) detectors. In the case of threshold detectors, we
have included as well the 
effect that detection inefficiencies and dark counts present in current measurement devices 
might have on the final secret ket rate. For simplicity, 
these measurement imperfections  were not considered in the original proposal.
On the other hand, PNR detectors have allowed us 
to obtain ultimate 
lower bounds on the maximal performance that can be expected at all from this kind of passive setups. 
We have also presented a passive scheme that employs
strong coherent light and does not require the use of  
single photon detectors, but it can operate with a simpler
classical photo-detector. This fact makes this setup specially interesting
from an experimental point of view. Finally, we have provided an estimation on the effects 
that statistical fluctuations due to a finite data size can have in practical implementations. 

\section{Acknowledgements}

The authors wish to thank H.-K. Lo, N. L\"utkenhaus, and
Y. Zhao for very useful discussions, and in particular M. Koashi for pointing out a reference.
M.C. especially thanks the University of 
Toronto and the  
Institute for Quantum Computing (University of Waterloo) for hospitality and support
during his stay in both institutions. This work was supported by the European Projects 
SECOQC and QAP, by the NSERC Discovery Grant, Quantum Works, CSEC, and
by Xunta de Galicia (Spain, Grant No. INCITE08PXIB322257PR). 

\appendix

\section{Estimation procedure}\label{ap_vico1}

Our starting point is the secret key rate formula given by Eq.~(\ref{ind_kr_new}). This expression 
can be lower bounded by
\begin{eqnarray}\label{ind_kr}
R^{l}&\geq&q\{-Q^{l}f(E^{l})H(E^{l})+(p^{l}_{1}Y_1+p^{l}_{0}Y_0)  \nonumber \\
&\times&[1-H(e_1^U)]\},
\end{eqnarray}
where $e_1^U$ denotes an upper bound on the single photon error rate $e_1$.
Hence, for our purposes it is enough 
to obtain a lower bound on the quantities
$p^{l}_{1}Y_1+p^{l}_{0}Y_0$ for all $l$, together with $e_1^U$. 
For that, we follow the estimation procedure proposed in Ref.~\cite{estimation}. 
Next, we show the explicit calculations for the case where Alice uses the passive scheme 
introduced in Sec.~\ref{sec_thermal}. 

\subsection{Lower bound on $p^{l}_{1}Y_1+p^{l}_{0}Y_0$}

The method contains two main steps. 
First, we have that $p^{l}_{1}Y_1+p^{l}_{0}Y_0$ always satisfies
\begin{equation}
p^{l}_{1}Y_1+p^{l}_{0}Y_0\geq{}p^{l}_{1}Y_1^L+p^{l}_{0}Y_0, 
\end{equation}
for all $l\in\{c,\bar{c}\}$, and
where $Y_1^L$ denotes a lower bound on the yield of a single photon state. To find 
$Y_1^L$, note that
\begin{eqnarray}\label{q1}
p^t_2Q^{\bar{c}}-p^{\bar{c}}_2Q^t&=&\sum_{n=0}^\infty (p^t_2p^{\bar{c}}_n-p^{\bar{c}}_2p^t_n)Y_n \nonumber \\
&\leq{}&\sum_{n=0}^1 (p^t_2p^{\bar{c}}_n-p^{\bar{c}}_2p^t_n)Y_n,
\end{eqnarray}
since 
\begin{eqnarray}\label{a2}
p^t_2p^{\bar{c}}_n-p^{\bar{c}}_2p^t_n&=&\frac{(1-\epsilon)(\mu{}t)^{n+2}}{[(1+\mu{}t)r]^3}\Big(\frac{1}{r^{n-2}} \nonumber \\
&-&\frac{1}{(1+\mu{}t)^{n-2}}\Big)\leq{}0,
\end{eqnarray}
for all $n\geq{}2$, and where the parameter 
$r$ is given by Eq.~(\ref{parameter_r}). To see this, 
note that the first term on the r.h.s. of Eq.~(\ref{a2}) is always greater or equal than zero, 
and $r\geq{}1+\mu{}t\geq{}1$. 
Similarly, we have that $p^t_2p^{\bar{c}}_n-p^{\bar{c}}_2p^t_n\geq{}0$ for all $n\leq{}1$.
Combining both results, we obtain
\begin{equation}\label{eq1_thermal}
Y_1\geq{}Y^L_1=\textrm{max}\Bigg\{\frac{p^t_2Q^{\bar{c}}-p^{\bar{c}}_2Q^t - (p^t_2p^{\bar{c}}_0-p^{\bar{c}}_2p^t_0)Y_0} {p^t_2p^{\bar{c}}_1-p^{\bar{c}}_2p^t_1},0\Bigg\}.
\end{equation}
Now comes the second step. The term which multiplies $Y_0$ in the 
expression
$p^{l}_{1}Y_1^L+p^{l}_{0}Y_0$ satisfies
\begin{equation}\label{rome2}
-p^{l}_{1}\frac{p^t_2p^{\bar{c}}_0-p^{\bar{c}}_2p^t_0} {p^t_2p^{\bar{c}}_1-p^{\bar{c}}_2p^t_1}+p^{l}_{0}\leq{}0.
\end{equation}
This last statement can be proven as follows. The condition given by Eq.~(\ref{rome2})
is equivalent to
\begin{equation}\label{aux_rome}
p^{l}_{0}(p^t_2p^{\bar{c}}_1-p^{\bar{c}}_2p^t_1)\leq{}p^{l}_{1}(p^t_2p^{\bar{c}}_0-p^{\bar{c}}_2p^t_0),
\end{equation}
since, as we have seen above, $p^t_2p^{\bar{c}}_1-p^{\bar{c}}_2p^t_1\geq{}0$. 
After a short calculation, 
it turns out that Eq.~(\ref{aux_rome}) can be further simplified to  
\begin{equation}
p^t_1p^{\bar{c}}_0-p^{\bar{c}}_1p^t_0\geq{}0, 
\end{equation}
both for $l=c$ and $l=\bar{c}$. Finally, from the definition of the probabilities $p^t_n$ and $p^{\bar{c}}_n$
given by Eqs.~(\ref{help_thermal})-(\ref{pnc_thermal_imp}), we find that
\begin{eqnarray}\label{q7}
p^t_1p^{\bar{c}}_n-p^{\bar{c}}_1p^t_n&=&\frac{(1-\epsilon)(\mu{}t)^{n+1}}{[(1+\mu{}t)r]^2} \nonumber \\
&\times&\Bigg(\frac{1}{r^{n-1}}-
\frac{1}{(1+\mu{}t)^{n-1}}\Bigg), 
\end{eqnarray}
which is greater or equal than zero 
for all $n\leq{}1$, and negative otherwise. Note that 
the first term on the r.h.s. of Eq.~(\ref{q7}) is always greater or equal than zero, and the sign of 
the second term depends on the value of $n$, since $r\geq{}1+\mu{}t\geq{}1$.

We obtain, therefore, that 
\begin{eqnarray}\label{rome_tues1}
p^{l}_{1}Y_1+p^{l}_{0}Y_0&\geq{}&\textrm{max}\Bigg\{\frac{p^{l}_{1}(p^t_2Q^{\bar{c}}-p^{\bar{c}}_2Q^t)}{p^t_2p^{\bar{c}}_1-p^{\bar{c}}_2p^t_1}  \\
&+&\Bigg[p^{l}_{0}
-p^{l}_{1}\frac{p^t_2p^{\bar{c}}_0-p^{\bar{c}}_2p^t_0} {p^t_2p^{\bar{c}}_1-p^{\bar{c}}_2p^t_1}\Bigg]Y_0^u,0\Bigg\},
\nonumber
\end{eqnarray}
for all $l\in\{c,\bar{c}\}$, and
where
$Y_0^u$ denotes an upper bound on the background rate $Y_0$. This parameter can be calculated from 
Eq.(\ref{rome3}). In particular, we have that
\begin{equation}\label{rome4}
Q^{c}E^{c}=\sum_{n=0}^\infty p^{c}_{n}Y_ne_n\geq{}p^{c}_{0}Y_0e_0,
\end{equation}
and similarly for the product $Q^{\bar{c}}E^{\bar{c}}$. We find
\begin{equation}
Y_0\leq{}Y_0^u=\textrm{min}\Bigg\{\frac{E^{c}Q^{c}}{p^{c}_0e_0},\frac{E^{\bar{c}}Q^{\bar{c}}}{p^{\bar{c}}_0e_0}\Bigg\}.
\end{equation}

\subsection{Upper bound on $e_1$} 

For this, we proceed 
as follows:
\begin{eqnarray}\label{q5}
p^{\bar{c}}_{0}Q^{t}E^{t}-p^{t}_{0}Q^{\bar{c}}E^{\bar{c}}&=&\sum_{n=1}^\infty 
(p^{t}_{n}p^{\bar{c}}_{0}-p^{\bar{c}}_{n}p^{t}_{0})Y_ne_n \nonumber \\
&\geq&(p^{t}_{1}p^{\bar{c}}_{0}-p^{\bar{c}}_{1}p^{t}_{0})Y_1e_1,
\end{eqnarray}
where the inequality condition comes from the fact that 
\begin{equation}\label{q4}
p^{t}_{n}p^{\bar{c}}_{0}-p^{\bar{c}}_{n}p^{t}_{0}=\frac{(1-\epsilon)(\mu{}t)^n}{(1+\mu{}t)r}
\Bigg(\frac{1}{(1+\mu{}t)^n}-\frac{1}{r^n}\Bigg)\geq{}0,
\end{equation}
for all $n\geq{}1$. 
From Eq.~(\ref{q5}) we obtain, therefore, that $e_1$ is upper bounded by
$(p^{\bar{c}}_0E^tQ^t
 -p^t_0E^{\bar{c}}Q^{\bar{c}})/[(p^t_1p^{\bar{c}}_0-p^{\bar{c}}_1p^t_0)Y_1^L]$, 
 where $Y_1^L$ is
 given by Eq.~(\ref{eq1_thermal}) with the parameter $Y_0$ replaced by $Y_0^u$. 

On the other hand, note that Eq.(\ref{rome3}) also provides a simple upper bound on $e_1$. Specifically, 
\begin{equation}
Q^{c}E^{c}=\sum_{n=0}^\infty p^{c}_{n}Y_ne_n\geq{}p^{c}_{0}Y_0e_0+p^{c}_{1}Y_1e_1, 
\end{equation}
and similarly for the product $Q^{\bar{c}}E^{\bar{c}}$. Putting all these conditions together, we find that
\begin{eqnarray}\label{eq2}
e_1\leq{}e^U_1=&\textrm{min}&\Bigg\{
\frac{E^{c}Q^{c}-p^{c}_0Y_0^Le_0}
{p^{c}_1Y_1^L}, 
\frac{E^{\bar{c}}Q^{\bar{c}}-p^{\bar{c}}_0Y_0^Le_0}{p^{\bar{c}}_1Y_1^L}, \nonumber \\
&& \frac{p^{\bar{c}}_0E^tQ^t
 -p^t_0E^{\bar{c}}Q^{\bar{c}}} 
 {(p^t_1p^{\bar{c}}_0-p^{\bar{c}}_1p^t_0)Y_1^L}
 \Bigg\},
\end{eqnarray} 
where 
$Y_0^L$ represents a lower bound on the background rate $Y_0$. To calculate this parameter we use the
following inequality: 
\begin{eqnarray}\label{q8}
p^t_1Q^{\bar{c}}-p^{\bar{c}}_1Q^t&=&(p^t_1p^{\bar{c}}_0-p^{\bar{c}}_1p^t_0)Y_0+\sum_{n=2}^\infty 
(p^t_1p^{\bar{c}}_n-p^{\bar{c}}_1p^t_n)Y_n \nonumber \\
&\leq&(p^t_1p^{\bar{c}}_0-p^{\bar{c}}_1p^t_0)Y_0,
\end{eqnarray}
since, as we have seen above, 
$p^t_1p^{\bar{c}}_n-p^{\bar{c}}_1p^t_n\leq{}0$
for all $n\geq{}2$. 
From Eq.~(\ref{q8}) we obtain, therefore, that
\begin{equation}\label{q6}
Y_0\geq{}Y^L_0=\textrm{max}\Bigg\{\frac{p^t_1Q^{\bar{c}}-p^{\bar{c}}_1Q^t}{p^t_1p^{\bar{c}}_0-p^{\bar{c}}_1p^t_0},0
\Bigg\}.
\end{equation}

\section{PNR detector}

In this Appendix we study the case where 
Alice uses a {\it perfect} PNR detector to measure the signal states in mode $b$. The main goal of 
this analysis is to obtain an ultimate 
lower bound on the secret key rate that can be achieved at all with the passive decoy state setups
introduced in Sec.~\ref{sec_thermal} and Sec.~\ref{sec_weak}, in combination with the 
security analysis provided in Refs.~\cite{gllp,lo_qic}. 

A perfect PNR detector can be characterized by a POVM which contains an infinite number of
elements, 
\begin{equation}\label{rome_cans3}
F_m=\ket{m}\bra{m},  
\end{equation}
with $m=0,1, \dots, \infty$. The outcome of $F_m$ corresponds to the detection of $m$ photons in mode $b$. 

\subsection{Thermal light}\label{ap_pnr} 

\begin{figure}
\begin{center}
\includegraphics[angle=0,scale=0.72]{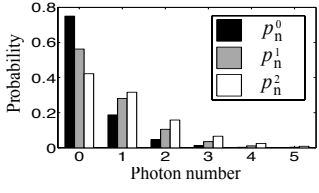}
\end{center}
\caption{
Conditional photon number distribution in mode $a$ when Alice uses a PNR detector, 
$\rho$ is given by Eq.~(\ref{rome_ns}), and $\sigma$ is a vacuum state: 
$p^{0}_{n}$ (black), $p^{1}_{n}$ (grey), and $p^{2}_{n}$ (white). We consider that $\mu=1$, $t=1/2$, and $n\leq{}5$.
\label{figure2_thermal_pnr}}
\end{figure}

Let us begin by considering the passive scheme analyzed in Sec.~\ref{sec_thermal} with 
Alice using a PNR detector. 
Whenever she finds $m$ photons in mode $b$, then the joint probability 
distribution of having $n$ photons in mode $a$ is just given by
Eq.~(\ref{prob_th}). 
Figure~\ref{figure2_thermal_pnr} shows the conditional 
photon number statistics in mode $a$ given that mode $b$ contains exactly
$m$ photons:
$p_n^m=p_{n,m}/N_m$, with 
\begin{equation}
N_m=\sum_{n=0}^{\infty} p_{n,m}=\frac{1}{1+\mu{}(1-t)}\Bigg[\frac{\mu{}(1-t)}{1+\mu{}(1-t)}\Bigg]^m.
\end{equation} 

In this scenario, it turns out that 
Alice and Bob can always estimate any finite number of yields $Y_n$ and error rates 
$e_n$ 
with arbitrary precision. In particular, they can obtain the actual values of the parameters
$Y_0$, $Y_1$, and $e_1$.
To see this, let $Q^m$ denote the overall gain of the signal states sent 
to Bob when mode $b$ contains exactly $m$ photons, and let the parameters $X_m$ and $V_n$ be defined as
\begin{eqnarray}\label{rome_monday}
X_m&=&\frac{(1+\mu)^{m+1}Q^m}{[\mu{}(1-t)]^m}, \nonumber \\
V_n&=&\Bigg[\frac{\mu{}t}{1+\mu}\Bigg]^nY_n.
\end{eqnarray}
With this notation, and using the 
definition of $p_{n,m}$ given by
Eq.~(\ref{prob_th}), we find that 
Eq.~(\ref{gain_rome}) can be rewritten as
\begin{equation}\label{rome_late}
X_m=\sum_{n=0}^\infty \binom{n+m}{m} V_n.
\end{equation}
That is,  
the coefficient matrix of the system of linear equations given by Eq.~(\ref{rome_late}) for all 
possible values of $m$
is a symmetric Pascal matrix \cite{matrix}. This matrix has determinant equal to one
and, therefore, in principle can always be inverted \cite{matrix}. 
Then, from the knowledge of the coefficients $V_n$, the legitimate users 
can directly obtain the values of the yields  
$Y_n$ by means of Eq.~(\ref{rome_monday}).
A similar argument can also be used to show that Alice and Bob can 
 obtain as well the values of $e_n$. 

After substituting Eqs.~(\ref{yield_new})-(\ref{qber_new})
into the gain and QBER formulas we obtain
\begin{eqnarray}
Q^{m}&=&N_m-\frac{(1-Y_0)[\mu(1-t)]^m}{\{1+\mu[1-(1-\eta_{\rm sys})t]\}^{m+1}}, \nonumber \\
Q^{m}E^{m}&=&(e_0-e_d)Y_0N_m+e_dQ^{m}.
\end{eqnarray}
In order to evaluate Eq.~(\ref{ind_kr_new})
we need to find the probabilities $p_{0,m}$ and $p_{1,m}$ for all $m$. From Eq.~(\ref{prob_th}) we have 
that these parameters can be expressed as
\begin{eqnarray}
p_{0,m}&=&\frac{[\mu(1-t)]^m}{(1+\mu)^{m+1}}, \nonumber \\
p_{1,m}&=&\frac{(m+1)t(1-t)^m}{1+\mu}\Big(\frac{\mu}{1+\mu}\Big)^{m+1}.
\end{eqnarray}

The resulting lower bound on the 
secret key rate is illustrated in Fig.~\ref{figure3_thermal_combined} (solid line).
The optimal values 
of the parameters $\mu$ and $t$ are quite constant with the 
distance. Specifically, in this figure
we choose $\mu$ around $18.5$ and $t$ around $0.02$.

\subsection{Weak coherent light}\label{ap_pnr2}

\begin{figure}
\begin{center}
\includegraphics[angle=0,scale=0.72]{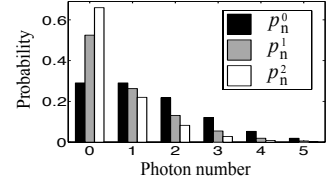}
\end{center}
\caption{
Conditional photon number distribution in mode $a$ when Alice uses a PNR detector: 
$p^{0}_{n}$ (black), $p^{1}_{n}$ (grey), and $p^{2}_{n}$ (white). 
The signal states $\rho$ and $\sigma$ in Fig.~\ref{figure_general} are two phase randomized WCP given by 
Eq.~(\ref{rome_cans2}).
We consider that $\mu_1=\mu_2=1$, $t=1/2$, and $n\leq{}5$.
\label{figure2_wcp_pnr}}
\end{figure}

Let us now consider the passive scheme illustrated in Sec.~\ref{sec_weak} with Alice 
using a PNR detector. Whenever her detector finds $m$ photons
in mode $b$, the joint probability distribution of having $n$ photons in mode $a$ is given by
Eq.~(\ref{q10}).
Figure~\ref{figure2_wcp_pnr} shows the conditional 
photon number statistics in mode $a$ given that mode $b$ contains exactly
$m$ photons: $p_n^m=p_{n,m}/N_m$, with  
\begin{equation}
N_m=\sum_{n=0}^\infty p_{n,m}=\frac{\upsilon^{m}e^{-\upsilon}}{m!}\frac{1}{2\pi}\int_0^{2\pi}(1-\gamma)^m e^{\upsilon\gamma} d\theta.
\end{equation}

To show that the experimental observations associated to different outcomes 
of the PNR detector allow Alice and Bob to obtain the 
values of the parameters $Y_0$, $Y_1$, and $e_1$ 
with arbitrary precision, one 
could follow the same procedure explained in Appendix~\ref{ap_pnr}. That is,  
one could try to prove that the determinant of the coefficient matrices associated to the systems of 
linear equations 
given by Eqs.~(\ref{gain_rome})-(\ref{rome3}) is different from zero also in this 
scenario. For simplicity, here we have confirmed this statement only 
numerically.

After substituting Eqs.~(\ref{yield_new})-(\ref{qber_new})
into the gain and QBER formulas we obtain
\begin{eqnarray}
Q^{m}&=&\frac{\upsilon^{m}e^{-\upsilon}}{m!}\frac{1}{2\pi}\int_0^{2\pi}[1-(1-Y_0)e^{-\eta_{\rm sys}\upsilon\gamma}] \nonumber \\
&\times&(1-\gamma)^m e^{\upsilon\gamma} d\theta, \nonumber \\
Q^{m}E^{m}&=&(e_0-e_d)Y_0N_m+e_dQ^{m}.
\end{eqnarray}
The relevant probabilities $p_{0,m}$ and $p_{1,m}$ can be 
calculated directly from Eq.~(\ref{q10}). We find that
\begin{eqnarray}\label{rome_bar}
p_{0,m}&=&\frac{e^{-\upsilon}(\upsilon-\omega)^m}{\Gamma_{1+m}}g
\Big[\frac{1-m}{2},-\frac{m}{2},1,\frac{\xi^2}{(\upsilon-\omega)^2}\Big], \nonumber \\
p_{1,m}&=&\omega{}p_{0,m}-\frac{e^{-\upsilon}\xi^2(\upsilon-\omega)^{m-1}}{2\Gamma_{m}} \nonumber \\
&\times&g\Big[\frac{1-m}{2},1-\frac{m}{2},2,\frac{\xi^2}{(\upsilon-\omega)^2}\Big], 
\end{eqnarray}
where the Gamma function $\Gamma_z$ is defined as \cite{Bessel}
\begin{equation}
\Gamma_z=\int_{0}^{\infty} t^{z-1}e^{-t} dt, 
\end{equation}
and 
where $g(a,b,c,z)$ represents the hypergeometric function \cite{Bessel}. This function is defined as \cite{Bessel}
\begin{equation}
g(a,b,c,z)=\frac{\Gamma_c}{\Gamma_b\Gamma_{c-b}}\int_{0}^1 \frac{t^{b-1}(1-t)^{c-b-1}}{(1-tz)^{a}} dt.
\end{equation}

In this case, the lower bound on the resulting secret key rate 
reproduces approximately the behavior of the asymptotic
active decoy state setup illustrated in Fig.~\ref{figure3} (solid line). 
Here we have assumed again that  
$t=1/2$. The values of the intensities $\mu_1$ and $\mu_2$ which optimize 
the secret key rate formula are, respectively, $\approx10^{-4}$ and  
$\approx0.95$. As already discussed in Sec.~\ref{sec_weak}, this result is not surprising since
the only difference between both setups (passive and active) arises from the
photon number probabilities of the signal states sent by Alice. 
While in the passive scheme the relevant statistics are given by Eq.~(\ref{rome_bar}), 
in the active setup they have the form 
\begin{eqnarray}\label{nap1}
p_{0,m}&=&e^{-\mu_m}, \nonumber \\
p_{1,m}&=&e^{-\mu_m}\mu_m,
\end{eqnarray}
with $\mu_m$ denoting the mean photon number of the signals associated to 
setting $m$. Still, it turns out that this difference is not significant enough to
be appreciated with the resolution of Fig.~\ref{figure3} when we optimize the parameters 
$\mu_1$ and $\mu_2$.

\section{Weak coherent light: Probabilities $p^{t}_n$ and $p^{\bar{c}}_n$}\label{apA}

In this Appendix we provide explicit expressions for the probabilities $p^{t}_n$ and $p^{\bar{c}}_n$, with $n=0,1,2$,
for the 
case of a passive decoy state setup with phase randomized WCP. After a short calculation, we find that
\begin{eqnarray}
p^{t}_0&=&I_{0,\xi}e^{-\omega}, \nonumber \\
p^{t}_1&=&(\omega{}I_{0,\xi}-\xi{}I_{1,\xi})e^{-\omega}, \nonumber \\
p^{t}_2&=&\frac{1}{2}\Big[\omega^2I_{0,\xi}+(1-2\omega)\xi{}I_{1,\xi}+\xi^2{}I_{2,\xi}\Big]e^{-\omega},
\end{eqnarray}
with 
$\omega=\mu_1{}t+\mu_2(1-t)$. The probabilities 
$p^{\bar{c}}_n$ have the form 
\begin{eqnarray}
p^{\bar{c}}_0&=&\tau{}I_{0,(1-\eta_{\rm d})\xi}, \nonumber \\
p^{\bar{c}}_1&=&\tau(\omega{}I_{0,(1-\eta_{\rm d})\xi}-\xi{}I_{1,(1-\eta_{\rm d})\xi}), \nonumber \\
p^{\bar{c}}_2&=&\frac{\tau}{2}\Big\{\omega^2I_{0,(1-\eta_{\rm d})\xi}+\Big[\frac{1}{1-\eta_{\rm d}}-2\omega\Big]\xi{}I_{1,(1-\eta_{\rm d})\xi} \nonumber \\
&+&\xi^2I_{2,(1-\eta_{\rm d})\xi}\Big\},
\end{eqnarray}
where $\tau=(1-\epsilon)e^{-[\eta_{\rm d}\upsilon+(1-\eta_{\rm d})\omega]}$. 

\section{Probabilities $p_n^{<I_M}$ and $p_n^{>I_M}$}\label{apB}

In this Appendix we provide explicit expressions for the probabilities $p_n^{<I_M}$ and $p_n^{>I_M}$, with $n=0,1,2$. 
For simplicity, we impose $I_1=I_2=I_M\equiv{}I$. 
After a short calculation, we obtain
\begin{eqnarray}
p^{<I_M}_0&=&\frac{e^{-\kappa}}{2}(I_{0,\zeta}-L_{0,\zeta}),  \\
p^{<I_M}_1&=&\frac{e^{-\kappa}}{2}[\kappa(I_{0,\zeta}-L_{0,\zeta})-\zeta(I_{1,\zeta}-L_{-1,\zeta})], \nonumber \\
p^{<I_M}_2&=&\frac{e^{-\kappa}}{4}\Big\{\kappa^2(I_{0,\zeta}-L_{0,\zeta})+\zeta\Big[\frac{2}{\pi}\Big(1-\frac{\zeta^2}{3}\Big) \nonumber \\
&+&(1-2\kappa)(I_{1,\zeta}-L_{-1,\zeta})+\zeta(I_{2,\zeta}-L_{2,\zeta})\Big]\Big\}, \nonumber
\end{eqnarray}
where $\kappa=It_2$, 
$\zeta=2\kappa\sqrt{t_1r_1}$, and
$L_{q,z}$ represents the modified Struve function
\cite{Bessel2}. 
This function is defined as \cite{Bessel2}
\begin{equation}\label{struve}
L_{q,z}=\frac{z^q}{2^{q-1}\sqrt{\pi}\Gamma_{q+1/2}}\int_{0}^{\pi/2} \sinh{(z\cos{\theta})
\sin{\theta}^{2q}}d\theta. 
\end{equation}
On the other hand, the probabilities 
$p^{>I_M}_n$ have the form 
\begin{eqnarray}
p^{>I_M}_0&=&\frac{e^{-\kappa}}{2}(I_{0,\zeta}+L_{0,\zeta}),  \\
p^{>I_M}_1&=&\frac{e^{-\kappa}}{2}[\kappa(I_{0,\zeta}+L_{0,\zeta})-\zeta(I_{1,\zeta}+L_{-1,\zeta})], \nonumber \\
p^{>I_M}_2&=&\frac{e^{-\kappa}}{4}\Big\{\kappa^2(I_{0,\zeta}+L_{0,\zeta})+\zeta\Big[-\frac{2}{\pi}\Big(1-\frac{\zeta^2}{3}\Big) \nonumber \\
&+&(1-2\kappa)(I_{1,\zeta}+L_{-1,\zeta})+\zeta(I_{2,\zeta}+L_{2,\zeta})\Big]\Big\}. \nonumber
\end{eqnarray}

\bibliographystyle{apsrev}

\end{document}